%% file: main.tex
\documentclass[a4paper,UKenglish,cleveref, autoref]{lipics-v2019}
\usepackage{microtype}
\title{Timed Systems through the Lens of Logic}

\author{S. Akshay}{Department of CSE, IIT Bombay, India}{akshayss@cse.iitb.ac.in}{}{}
\author{Paul Gastin}{LSV, ENS Paris-Saclay \& CNRS, Universit{\'e} Paris-Saclay}{paul.gastin@ens-paris-saclay.fr}{}{}
\author{Vincent Jug\'e}{LIGM, Universit\'e Paris-Est Marne-la-Vall\'ee, CNRS}{vincent.juge@u-pem.fr}{}{}
\author{Shankara Narayanan Krishna}{Department of CSE, IIT Bombay, India}{krishnas@cse.iitb.ac.in}{}{}

\authorrunning{S. Akshay, P. Gastin, V. Jug\'e, S. Krishna}

\Copyright{S. Akshay, Paul Gastin, Vincent Jug\'e and Shankara Narayanan Krishna}



\funding{Partly supported by 
UMI ReLaX, 
ANR project TickTac (ANR-18-CE40-0015), 
DST/INRIA CEFIPRA project EQuaVe
and DST/INSPIRE faculty award [IFA12-MA-017].
}
\hideLIPIcs  

\input{preamble}

\begin{document}
\nolinenumbers
\maketitle

\begin{abstract}
In this paper, we analyze timed systems with data structures. 
 We start by describing behaviors
of timed systems using graphs with timing constraints. Such a graph is called
realizable if we can assign time-stamps to nodes or events so that they are
consistent with the timing constraints. The logical definability of several
graph properties~\cite{bruno1,bruno2} has been a challenging problem, and
we show, using a highly non-trivial argument, that the realizability property
for collections of graphs with strict timing constraints is logically definable
in a class of propositional dynamic logic (EQ-ICPDL), which is strictly
contained in MSO. Using this result, we propose a novel, algorithmically
efficient and uniform proof technique for the analysis of timed systems enriched
with auxiliary data structures, like stacks and queues. Our technique unravels
new results (for emptiness checking as well as model checking) for timed
systems with richer features than considered so far, while also recovering
existing results.
\end{abstract}

\section{Introduction}
\label{sec:intro}
The modeling and analysis of complex real-time systems is a challenging and
important area, both from theoretical and practical points of view.  The
challenge often stems from the fact that such models have different sources of
infinite behaviors, which makes them highly expressive but difficult to analyze.
On one hand, the timing features engender complex constraints between events,
which allow (or disallow) infinite sets of timed behaviors (over real numbers)
satisfying these constraints.  On the other hand, the auxiliary data structures
such as multiple stacks allow a rich expressive power often leading to
undecidable verification problems, even in the absence of time.  Thus, each
choice of combining these components of real-time and specific data structures
leads to rich models whose analysis is complicated and often intractable.

The analysis of timed systems without any additional data structures has often
been done using well-accepted models like timed automata~\cite{AD94}, where
clocks are real-valued variables that are reset and checked at guards.  The
classical approach to analyze such timed automata is by abstracting the
real-timed system using the so-called region abstraction into a finite-state
automaton preserving emptiness.  Several variants and extensions of this basic
model have been considered over the years, for instance using
event-clocks~\cite{AFH99} or diagonal constraints, or even by allowing
(non-)\,deterministic updates of clocks.  Subsequently, there has been a growing
body of
work~\cite{lics12,formats18,concur16,AGKS17,Clemente18,CL15,CL18,CLR17,hscc15}
towards adding auxiliary data structures like stacks~\cite{MP11,AGK14,CG14} or
queues~\cite{CG14} to such timed automata.  In all these, the techniques used to
solve the emptiness problem were specific and tailor-made to the choice of the data structure, kind of constraints and updates that are allowed.  
Our goal is to introduce a novel and uniform approach for reasoning about such
timed systems which allow rich timing features along with several types of
auxiliary data structures at the same time.  This technique captures the
behaviors of the underlying model as graphs (see~\cite{CG14}) and examines the
logical definability of certain properties over these graphs.

We start by abstracting a run of a system, be it timed or not, as a
sequence of instructions.  When the system has a data structure $d$ such as a
stack, these instructions may write to $d$ (denoted $w(d)$) or read from $d$ ($r(d)$).  The behavior is modeled as a linear graph (the sequence of instructions), with instruction labels and with additional data-structure edges matching writes with corresponding reads, as illustrated in Figure~\ref{fig:untimed-data}.  When the system is timed, instructions may also reset clocks ($x:=0$), check guards ($x<3$), etc.  These timing instructions are recorded as additional labels in the linear graph without a priori being interpreted as edges, as shown on Figure~\ref{fig:timed-data} left.  This allows to decouple the behavior of the underlying untimed system from the timing constraints that should be realized for the run to be feasible.

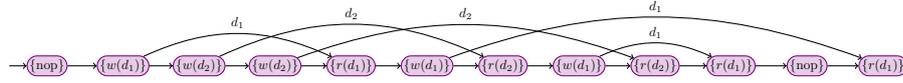
\begin{figure}[t]
 \begin{center}
 \scalebox{0.5}{
 \begin{tikzpicture}[->,thick]
 \node[initial, cir, initial text ={}] at (11,0) (aA) {$\{\nop\}$} ;

 \node[cir] at (13,0) (aB) {$\{w(d_1)\}$};

 \node[cir] at (15,0) (aC) {$\{w(d_2)\}$};

 \node[cir] at (17,0) (aD) {$\{w(d_2)\}$};

 \node[cir] at (19,0) (aE) {$\{r(d_1)\}$};

 \node[cir] at (21,0) (aF) {$\{w(d_1)\}$};%

 \node[cir] at (23,0) (aG) {$\{r(d_2)\}$};%

 \node[cir] at (25,0) (aH) {$\{w(d_1)\}$};%

 \node[cir] at (27,0) (aI) {$\{r(d_2)\}$};%

 \node[cir] at (29,0) (aJ) {$\{r(d_1)\}$};%

 \node[cir] at (31,0) (aK) {$\{\nop\}$};

 \node[cir] at (33,0) (aL) {$\{r(d_1)\}$};%

 \path (aA) edge node {}node {} (aB);
 \path (aB) edge node {}node {} (aC);
 \path (aC) edge node {}node {} (aD);
 \path (aD) edge node {}node {} (aE);
 \path (aE) edge node {}node {} (aF);
 \path (aF) edge node {}node {} (aG);
 \path (aG) edge node {}node {} (aH);
 \path (aH) edge node {}node {} (aI);
 \path (aI) edge node {}node {} (aJ);
 \path (aJ) edge node {}node {} (aK);
 \path (aK) edge node {}node {} (aL);

 \path(aB) edge[bend left=25] node[above] {$d_1$} node{}(aE);
 \path(aF) edge[bend left=20] node[above] {$d_1$} node{}(aL);
 \path(aH) edge[bend left=25] node[above] {$d_1$} node{}(aJ); 

 \path(aC) edge[bend left=25] node[above] {$d_2$} node{}(aG);
 \path(aD) edge[bend left=20] node[above] {$d_2$} node{}(aI); 

 \end{tikzpicture}
}
 \caption{Labeled linear graph $G_\sigma$ of a sequence of instructions $\sigma=\nop$ $~w(d_1)~$ $w(d_2)~$ $w(d_2)$ $~r(d_1)$ $~w(d_1)$ $~r(d_2)$ $~w(d_1) ~r(d_2)$ $~r(d_1)~\nop ~r(d_1)$ from a system having two data structures (a stack $d_1$ and a queue $d_2$).}
 \label{fig:untimed-data}
 \end{center}
\end{figure}
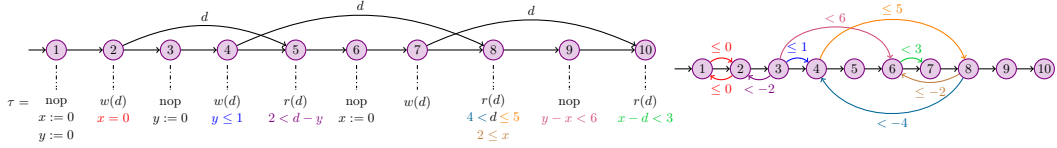
\begin{figure}[t]
 \begin{center}
 \scalebox{0.5}{
 \begin{tikzpicture}[->,thick] 

 \node[state, draw=white, rounded rectangle,minimum size=1.5em,inner sep=0em] at (-1,-1.4) (B0) {$\tau=$};

 \node[initial, cir, initial text ={}] at (0,0) (aA) {$1$} ;
 \draw[dashdotted,-] (0,-0.4)--(0,-1.1);
 \node[state, draw=white, rounded rectangle] at (0,-1.8) (B) {$\begin{array}{c} \nop \\ x:=0\\y:=0 \end{array}$};

 \node[cir] at (1.5,0) (aB) {$2$};
 \draw[dashdotted,-] (1.5,-0.4)--(1.5,-1.1);
 \node[state, draw=white, rounded rectangle] at (1.5,-1.6) (B1) {$\begin{array}{c} w(d) \\ \textcolor{red}{x=0} \\ \end{array}$};

 \node[cir] at (3,0) (aC) {$3$};
 \draw[dashdotted,-] (3,-0.4)--(3,-1.1);
 \node[state, draw=white, rounded rectangle] at (3,-1.6) (B2) {$\begin{array}{c} \nop \\ y:=0 \\ \end{array}$};

 \node[cir] at (4.5,0) (aD) {$4$};
 \draw[dashdotted,-] (4.5,-0.4)--(4.5,-1.1);
 \node[state, draw=white, rounded rectangle] at (4.5,-1.6) (B3) {$\begin{array}{c} w(d) \\ \textcolor{blue}{y \leq 1}\\ \end{array}$};

 \node[cir] at (6.3,0) (aE) {$5$};
 \draw[dashdotted,-] (6.3,-0.4)--(6.3,-1.1);
 \node[state, draw=white, rounded rectangle] at (6.3,-1.6) (B4) {$\begin{array}{c} r(d) \\ \textcolor{violet}{2<d-y} \end{array}$};

 \node[cir] at (7.9,0) (aF) {$6$};%
 \draw[dashdotted,-] (7.9,-0.4)--(7.9,-1.1);
 \node[state, draw=white, rounded rectangle] at (7.9,-1.6) (B5) {$\begin{array}{c} \nop \\ x:=0 \\ \end{array}$};

 \node[cir] at (9.5,0) (aG) {$7$};%
 \draw[dashdotted,-] (9.5,-0.4)--(9.5,-1.1);
 \node[state, draw=white, rounded rectangle] at (9.5,-1.4) (B6) {$\begin{array}{c} w(d) \end{array}$};

 \node[cir] at (11.5,0) (aH) {$8$};%
 \draw[dashdotted,-] (11.5,-0.4)--(11.5,-1.1);
 \node[state, draw=white, rounded rectangle] at (11.5,-1.8) (B7) {$\begin{array}{c} r(d) \\ \fole d \leqfi \\
 \textcolor{violet!50!yellow}{2 \leq x}\end{array}$};

 \node[cir] at (13.5,0) (aI) {$9$};%
 \draw[dashdotted,-] (13.5,-0.4)--(13.5,-1.1);
 \node[state, draw=white, rounded rectangle] at (13.5,-1.6) (B8) {$\begin{array}{c} \nop \\ \textcolor{purple!70!white}{y-x<6} \\ \end{array}$};

 \node[cir] at (15.5,0) (aJ) {$10$};%
 \draw[dashdotted,-] (15.5,-0.4)--(15.5,-1.1);
 \node[state, draw=white, rounded rectangle] at (15.5,-1.6) (B9) {$\begin{array}{c} r(d) \\ \textcolor{green!80!blue}{x-d<3} \\ \end{array}$};

 \path (aA) edge node {}node {} (aB);
 \path (aB) edge node {}node {} (aC);
 \path (aC) edge node {}node {} (aD);
 \path (aD) edge node {}node {} (aE);
 \path (aE) edge node {}node {} (aF);
 \path (aF) edge node {}node {} (aG);
 \path (aG) edge node {}node {} (aH);
 \path (aH) edge node {}node {} (aI);
 \path (aI) edge node {}node {} (aJ);

 \path(aB) edge[bend left=25] node[above] {$d$} node{}(aE);
 \path(aD) edge[bend left=25] node[above] {$d$} node{}(aH);
 \path(aG) edge[bend left=25] node[above] {$d$} node{}(aJ); 

 \node[initial, cir, initial text ={}] at (17,-0.5) (xA) {$1$} ;
 \node[cir] at (18,-0.5) (xB) {$2$};
 \node[cir] at (19,-0.5) (xC) {$3$};
 \node[cir] at (20,-0.5) (xD) {$4$};
 \node[cir] at (21,-0.5) (xE) {$5$};
 \node[cir] at (22,-0.5) (xF) {$6$};%
 \node[cir] at (23,-0.5) (xG) {$7$};%
 \node[cir] at (24,-0.5) (xH) {$8$};%
 \node[cir] at (25,-0.5) (xI) {$9$};%
 \node[cir] at (26,-0.5) (xJ) {$10$};%

 \path (xA) edge node {}node {} (xB);
 \path (xB) edge node {}node {} (xC);
 \path (xC) edge node {}node {} (xD);
 \path (xD) edge node {}node {} (xE);
 \path (xE) edge node {}node {} (xF);
 \path (xF) edge node {}node {} (xG);
 \path (xG) edge node {}node {} (xH);
 \path (xH) edge node {}node {} (xI);
 \path (xI) edge node {}node {} (xJ);

 \path(xA) edge[draw=red,bend left=40] node[above] {\textcolor{red}{$\leq 0$}} node{}(xB);
 \path(xB) edge[draw=red,bend left=40] node[below] {\textcolor{red}{$\leq 0$}} node{}(xA);
 \path(xC) edge[draw=violet,bend left=40] node[below] {\textcolor{violet}{$< -2$}} node{}(xB);
 \path(xC) edge[draw=blue,bend left=40] node[above] {\textcolor{blue}{$\leq 1$}} node{}(xD);
 \path(xC) edge[draw=purple!70!white,bend left=80] node[above] {\textcolor{purple!70!white}{$< 6$}} node{}(xF);

 \path(xD) edge[draw=red!50!yellow,bend left=70] node[above] {\textcolor{red!50!yellow}{$\leq 5$}} node{}(xH);
 \path(xH) edge[draw=blue!60!green,bend left=60] node[below] {\textcolor{blue!60!green}{$< -4$}} node{}(xD);

 \path(xF) edge[draw=green!80!blue, bend left=40] node[above] {\textcolor{green!80!blue}{$< 3$}} node{}(xG);
 \path(xH) edge[draw=violet!50!yellow,bend left=30] node[below] {\textcolor{violet!50!yellow}{$\leq -2$}} node{}(xF);

 \end{tikzpicture}
}
 \caption{Left: labeled linear graph $G_{\tau}$ obtained from a sequence of timed instructions $\tau$.
   For readability, the nodes are numbered and their instruction labels are written below them. Right: the corresponding weighted graph $\wg_\tau$.}
 \label{fig:timed-data}
 \end{center}
\end{figure}

 Our first contribution is to show that non-emptiness of a timed system $\TS$ can
  be reduced to the satisfiability of a formula $\Phi_\TS$ over such labeled linear graphs, which we call $\TS$-graphs. 
  A $\TS$-graph $G_\tau$ obtained from a sequence of instructions $\tau$, as depicted in Figure~\ref{fig:timed-data} (left), is a witness of non-emptiness of $\TS$ if it satisfies three properties:

\noindent 1) The sequence of instructions $\tau$ can be generated by $\TS$. 
Since the system $\TS$ is usually described with
a finite automaton where transitions are labeled with instructions, $\TS$ induces a \emph{regular} language of instruction sequences which can  easily be captured by ($\Phi_1$) in our logic.

\noindent 2) The data-structure edges should comply with the sequence of
instructions.  Intuitively, a node labeled with $w(d)$ (resp.\ $r(d)$) should
have an outgoing (resp.\ incoming) $d$-edge.  If the data structure $d$ is a stack (resp.\ queue), then $d$-edges should be well-nested, i.e., satisfy the LIFO
(resp.\ FIFO) policy.  It is known that compliance with stack or queue data-structures can be expressed ($\Phi_2$) in our logics~\cite{mpri-lect}.

\noindent 3) The real-time constraints induced by the timing instructions should be \emph{realizable}, i.e., it is possible to timestamp the nodes of $G$ with some real numbers so that all timing constraints are satisfied.  The second main contribution of this paper is to show that realizability can be expressed ($\Phi_3$) in our logic.

We use a light-weight propositional dynamic logic called
\pdl for the logical definability. Writing formulae for our systems in \pdl is
rather intuitive and improves readability in several cases compared to the
classical \mso. On a technical note, it is known that \pdl is a strict fragment
of \mso, and gives us a more tractable complexity than \mso{} (avoiding a
non-elementary blowup).

We show that realizability can be expressed in \pdl in two steps.  First, from
the $\TS$-graph $G_\tau$, we define a weighted graph $\wg_\tau$ which retains only the timing constraints induced by the timed instruction sequence $\tau$.  For
instance, in Figure~\ref{fig:timed-data}, the $\TS$-graph $G_\tau$ is on the
left and the associated weighted graph $\wg_\tau$ on the right.  In $\wg_\tau$,
an edge from node $i$ to node $j$ labeled $<6$ means that the difference
$t(j)-t(i)$ between the timestamps assigned to $i$ and $j$ should be less than
$6$.  We prove that the weighted graph $\wg_\tau$ can be \pdl-interpreted in the
graph $G_\tau$.  This holds for all timing features that we consider.  Second, we
prove that realizability of weighted graphs is expressible in \pdl, say with
$\Phi'_3$.  Since weighted graphs $\wg_\tau$ can be \pdl-interpreted in
$\TS$-graphs $G_\tau$, we can backward translate $\Phi'_3$ into some $\pdl$
formula $\Phi_3$ expressing realizability over $\TS$-graphs.  Finally,
non-emptiness of $\TS$ is equivalent to satisfiability of
$\Phi_\TS=\Phi_1\wedge\Phi_2\wedge\Phi_3$.

Our logical characterization of realizability for weighted graphs is highly
non-trivial. It is easier when the underlying system only has closed guards,
but we go beyond this and prove that realizability is also definable in \pdl in
the presence of both open and closed guards.  On the other hand, we show that,
without the linear order, realizability is not definable in \mso.  In fact, we
show that this already holds for graphs with a partial order of width (i.e.,
size of the largest anti-chain) 2, thus proving a tight characterization.

Our third contribution is to show how the two results above can be combined with
\emph{existing} techniques to give an effective algorithm for checking emptiness
of several classes of timed systems.  First, observe that the above two
contributions do not immediately imply that checking emptiness of the system is
decidable, as satisfiability of \pdl formulae over arbitrary collections of
graphs is undecidable.  This is expected, since, even in the untimed case, having
a single queue or two stacks as data structures leads to undecidability of
emptiness.  However, we can now consider under-approximations, as classically
done for untimed systems.  One such under-approximation is to consider
collections of $\TS$-graphs that have a fixed bound on the tree-width.  Such
$\TS$-graphs can now be interpreted into trees and we can use the fact that
checking satisfiability for \pdl (with bounded intersection width) over trees is
decidable in $\mathsf{EXPTIME}$.  This gives us a matching $\mathsf{EXPTIME}$
algorithm for checking emptiness of timed systems whose graph behaviors
have a bounded tree-width. 
Using this approach, we retrieve many known results on timed systems
with data structures, and also obtain new results.
Our approach captures with elan, the intricate flow and exchange of information between data structures and clocks, see Section~\ref{sec:ext}.

\subparagraph{Related work} Our technique is orthogonal to the theory of timed
systems via the region construction as well as to other related 
approaches.  In the untimed setting, the closest work to ours is
in~\cite{MP11,AGK14}, where generic approaches for decidability via logic and
tree-width have been developed for automata with data structures in the untimed
setting.  There have been several papers on the decidability of timed systems with
a single stack: \cite{BER94,lics12} deal with specific timing constraints,
while~\cite{CL15,CL18} use the language of timed atoms to specify and analyze an
orthogonal but powerful extension to timed registers.  In~\cite{CLR17}, a
$\mathsf{NEXPTIME}$ bound is shown in this setting by reduction to
one-dimensional branching vector addition systems.  However, all these works are
restricted to a single stack, while we tackle several data structures including
multiple stacks, queues.  Many recent papers \cite{CL18,Clemente18,formats18} consider complex constraints between data structures and clocks.
In these papers, there are time constraints between data structures $d_1, d_2$,
between clocks, and also between a clock $c$ and a data structure $d$.  All of
these can be modeled easily in our case, as can be seen in Section~\ref{sec:ext}.

Our work is also related to~\cite{concur16,AGKS17},
where the behaviors of timed systems with stacks are modeled as graphs
having data-structure edges as well as time constraint edges. 
The presence of two types of edges necessitates a fresh proof 
for the  the bound on tree-width for each kind of 
timing feature.  On the contrary, we directly 
inherit the bound on tree-width established in the untimed setting. 
The other main difference is that \cite{concur16,AGKS17} directly build tree automata instead of
going via logic.  Using logic instead of directly building a tree automaton
allows us to have a simpler higher level approach which is easier to write and
less technical.

The logic we use builds on Propositional Dynamic Logic, a classical logic to
reason about programs~\cite{FL79}.  The extension with loop, intersection and
converse was explored in~\cite{GLL09}, where complexity bounds were shown for
satisfiability and model checking.  We inherit these complexity bounds.
However, to the best of our knowledge, this is the first time this logic has
been used in the analysis of timed systems. Further, even with \mso logic (a strictly more powerful and
well-known logic), the characterization of realizability in \mso over graphs of
timed systems was open, as mentioned in~\cite{concur16}: we settle this problem
in this paper.

\section{Preliminaries}
\label{sec:prelim}

\subparagraph{Node- and edge-labeled graphs} %
Let $\Sigma$ and $\Gamma$ be two alphabets. Nodes will be labeled with
$\Sigma$ and edges with $\Gamma$. A $(\Sigma,\Gamma)$-labeled graph is a tuple
$G=(V,E,\lambda)$ where $V$ is a finite set of vertices, $\lambda\colon V\to
2^{\Sigma}$ labels vertices with (sets of) letters from $\Sigma$ and
$E\subseteq V\times\Gamma\times V$ is the set of labeled edges.
A vertex may have 0, 1 or several labels from $\Sigma$. For
$\gamma\in\Gamma$, we let $E_\gamma=\{(u,v)\myst(u,\gamma,v)\in E\}$ be the set
of edges labeled $\gamma$.
$\Graphs(\Sigma,\Gamma)$ denotes the set of $(\Sigma,\Gamma)$-labeled
graphs.

In this paper, graphs model behaviors of sequential systems. Hence, we have a
special symbol $\Succ$ in $\Gamma$ to define the successor relation $E_\Succ$ of
a total order on $V$. We simply write $u\podot v$ instead of $(u,v)\in
E_{\Succ}$. We call these graphs \emph{linear}; we let ${\po}={\podot}^{*}$
be the linear order induced by $\podot$ and we note ${\prec}={\podot}^{+}$ the
strict order. The other edges $E_\gamma$, with $\gamma \in \Gamma\setminus\{\Succ\}$, are
used to model other useful relations in the graph, for instance the matching
push-pop relation if we are interested in pushdown systems.

\subparagraph{Propositional dynamic logic over labeled graphs}
We define now the logic that we will use to specify properties of graphs. We
use a variant of the propositional dynamic logic~\cite{FL79}. This logic is sufficiently
expressive for our purposes and enjoys good complexity for the satisfiability
problem, rather than
the more expressive monadic second order logic
($\mathsf{MSO}$) which has a much higher complexity. The logic
$\ICPDL(\Sigma,\Gamma)$ is defined over $\Sigma$ (often seen as propositional
variables), and $\Gamma$ (often seen as atomic programs).

\subparagraph{Syntax}
We have the following, with $p\in\Sigma$ and $\gamma\in\Gamma$:
\begin{align*}
 \Phi &::= \existsnode \sigma \myst \neg \Phi \myst \Phi\vee\Phi\\
 \sigma &::= \top\myst p\myst \sigma\vee \sigma \myst \neg \sigma \myst
 \existspath{\pi} \sigma\myst \existsloop{\pi}\\
 \pi &::=
 {\Edge{\gamma}} \myst \test{\sigma}
 \myst \pi+\pi \myst \pi\cdot \pi\myst \pi^\ast \myst \pi^{-1} \myst \pi\cap\pi
\end{align*}
In $\ICPDL$, $\mathsf{C}$ stands for converse ($\pi^{-1}$) and $\mathsf{I}$ for
intersection ($\pi\cap\pi$). We also consider $\LCPDL$ which is the fragment
with loop but without intersection, since it has better complexity, as stated in
Theorem~\ref{thm:pdl-sat}. We also write $\CPDL$ or $\PDL$ with the
obvious meaning.
In the syntax above, $\Phi$ are sentences and $\existsnode$ is the existential node
quantifier. The universal node quantifier $\forallnodes\sigma$ is written
$\neg\existsnode\neg\sigma$. Formulae $\sigma$ are called \emph{node} or \emph{state}
formulae and have one
implicit free first-order variable, while formulae $\pi$ are called \emph{path} or \emph{program} formulae and have two implicit free first-order variables, the endpoints of the path.

\subparagraph{Semantics} Given a $(\Sigma,\Gamma)$-labeled graph
$G=(V,E,\lambda)$, we can write the semantics of the formulae.
The semantics of a \emph{state formula} $\sigma$ is a set
$\sem{\sigma}_{G}\subseteq V$, while the semantics of a \emph{path formula} $\pi$
is a binary relation $\sem{\pi}_{G} \subseteq V^{2}$. Their definitions are
mutually inductive. If the graph $G$ is clear from the context, we omit subscripts and simply write $\sem{\sigma}$ and $\sem{\pi}$.

The base cases for path formulae are $\sem{\Edge{\gamma}}=E_\gamma$ and
$\sem{\test{\sigma}}=\{(v,v)\myst v\in\sem{\sigma}\}$.
The operations $+, \cap, \cdot, ^\ast$ correspond to rational expression notations,
interpreted respectively as union, intersection, concatenation and Kleene star
of the respective relations. Finally, the converse is defined by
$\sem{\pi^{-1}}=\{(u,v)\myst(v,u)\in\sem{\pi}\}$.

The base cases for state formulae are $\sem{\top}=V$ and $\sem{p}=\{v\in V\myst
p\in \lambda (v)\}$, where $p\in\Sigma$. Disjunction and negation correspond to
union and complement. We let $\sem{\existsloop{\pi}}$ consist of the vertices
$v\in E$ from which there is a loop following path $\pi$, i.e., such that
$(v,v)\in\sem{\pi}$. Similarly, we let $\sem{\existspath{\pi}\sigma}$ consist
of the vertices $u\in E$ from which it is possible to follow the path $\pi$ and
reach a vertex satisfying $\sigma$, i.e., $(u,v)\in\sem{\pi}$ for some
$v\in\sem{\sigma}$. We often write $\existspath{\pi}$ instead of
$\existspath{\pi}\top$.
A sentence $\existsnode\sigma$ states that there exists a vertex of $G$
satisfying $\sigma$, i.e., $G\models\existsnode\sigma$ if
$\sem{\sigma}_{G}\neq\emptyset$. Disjunction and negation of sentences
are as usual. 

While $\ICPDL$ allows intersection, loop and converse, we also look at \pdl
where we allow existential quantification over new propositional variables in a
similar spirit as in~\cite{qctl}. Thus, formulae of $\pdl(\Sigma,\Gamma)$ have
the form $\Psi=\exists p_1,\ldots,p_n~\Phi$ where $\AP=\{p_1,\ldots,p_n\}$ is
disjoint from $\Sigma$ and $\Phi \in \ICPDL(\Sigma\uplus\AP,\Gamma)$.
The semantics is defined by $G=(V,E,\lambda)\models\exists
p_1,\ldots,p_n~\Phi$ if there exists $\lambda'\colon V\to 2^{\mathsf{AP}}$ such
that $(G,\lambda')=(V,E,\lambda\cup\lambda')\models\Phi$.
 For formulae $\Psi$ in $\ICPDL(\Sigma,\Gamma)$ or $\pdl(\Sigma,\Gamma)$, we
let $L(\Psi)=\{G\in\Graphs(\Sigma,\Gamma) \myst G \models\Psi\}$.
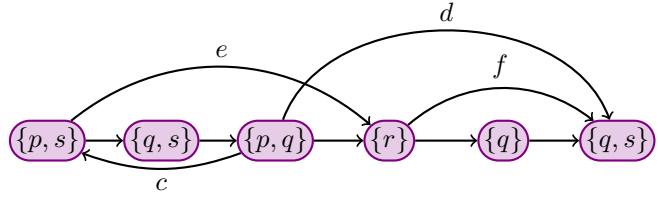
\begin{figure} [t] \begin{center}
 \begin{tikzpicture}[->,thick]
 \node[cir] at (-20,0) (A) {$\{p,s\}$} ;
 \node[cir] at (-18.5,0) (B) {$\{q,s\}$};
 \node[cir] at (-17,0) (C) {$\{p,q\}$};
 \node[cir] at (-15.5,0) (D) {$\{r\}$};
 \node[cir] at (-14,0) (E) {$\{q\}$};
 \node[cir] at (-12.5,0) (F) {$\{q,s\}$};

 \path (A) edge node {}node {} (B);
 \path (B) edge node {}node {} (C);
 \path (C) edge node {}node {} (D);
 \path (D) edge node {}node {} (E);
 \path (E) edge node {}node {} (F);

 \path(C) edge[bend left=70] node[above] {$d$} node{}(F);
 \path(A) edge[bend left=40] node[above] {$e$} node{}(D);
 \path(D) edge[bend left=40] node[above] {$f$} node{}(F);
 \path(C) edge[bend left=20] node[below] {$c$} node{}(A);
 \end{tikzpicture}
 \caption{A node- and edge-labeled graph.
 }
 \label{fig:icpdl}
\end{center}
\end{figure}

\begin{example}
 We illustrate the semantics of $\ICPDL(\Sigma,\Gamma)$ using Figure \ref{fig:icpdl}.
 We have a node- and edge-labeled graph, with node labels $\Sigma=\{p,q,r,s\}$
 and edge labels $\Gamma=\{d,e,f,\Succ\}$. In path formulae, we simply write
 $\rightarrow$ instead of $\Edge{\Succ}$. The formula $\existsnode
 \existspath{(\test{p \vee q}\cdot{\rightarrow})^\ast}r$ evaluates to true on the given
 graph: the 
 leftmost node is a witness.
 Likewise, the formula $\neg
 \existsnode \existspath{\rightarrow}(p \wedge s)$ is also true, since there
 are no nodes in the graph whose successors are labeled both $p$ and $s$. Let
 $\Delta=\Gamma \setminus \{\Succ\}$. The formula $\existsnode
 \bigvee_{(d,d') \in \Delta^2, d \neq d'}
 \existsloop{{\Edge{d}}\cdot{\Edgem{d'}}}$ is not true since all the
 non-successor edges are labeled by a unique symbol.
 Finally, the formula
 $\existsnode
 \existspath{\test{s}\cdot{\Edge{e}}\cdot\test{r}\cdot{\Edge{f}}
 \cdot\test{s}\cdot{\Edgem{d}}\cdot {\Edge{c}}}p$
 is true,
 while $\existsnode
 \existspath{\test{p}\cdot{\Edge{d}}}r$ is not.
\end{example}

\subparagraph{Satisfiability of propositional dynamic logic} %
The following definitions and results will be used in Section~\ref{sec:sat-complexity}.
Over arbitrary graphs, the satisfiability problem for \textsf{PDL} is
undecidable. On the other hand, when we restrict to graphs of bounded
tree-width, then the satisfiability problem becomes decidable with elementary
complexity. We explain this now. Tree-width is a well-known measure for graphs~\cite{RS84}. We say that a labeled graph
$G=(V,E,\lambda)$ has tree-width $k$ if the underlying unlabeled graph has
tree-width $k$. We will not need the formal definition of tree-width in this
paper, so it is omitted. We denote by $\Graphs^{k}(\Sigma,\Gamma)$ the graphs
in $\Graphs(\Sigma,\Gamma)$ having tree-width at most $k$.

Below is one of the main theorems that we use in this paper.
It refers to the intersection width of an $\EQICPDL$ formula, which is the
maximum of the intersection widths of its path subformulae: the intersection
width of path formulae is defined inductively by
$\mathsf{iw}(\Edge{\gamma})=\mathsf{iw}(\test{\sigma})=1$,
$\mathsf{iw}(\pi_1+\pi_2)=\mathsf{iw}(\pi_1\cdot\pi_2)=\max(\mathsf{iw}(\pi_1),\mathsf{iw}(\pi_2))$,
$\mathsf{iw}(\pi^{-1})=\mathsf{iw}(\pi^{*})=\mathsf{iw}(\pi)$, and
$\mathsf{iw}(\pi_1\cap\pi_2)=\mathsf{iw}(\pi_1)+\mathsf{iw}(\pi_2)$. Hence, a
formula in $\LCPDL$ has intersection width 1.

\begin{theorem}[Satisfiability]\label{thm:pdl-sat}
 Given $k\geq 1$ in unary and a formula $\Psi$ in $\EQICPDL(\Sigma,\Gamma)$
 of intersection width bounded by a constant, checking whether
 $G\models\Psi$ for some $G\in\Graphs^{k}(\Sigma,\Gamma)$ can be solved in
 $\mathsf{EXPTIME}$.
\end{theorem}

This is a consequence of a similar result over trees due to G\"oller, Lohrey and
Lutz \cite[Theorem 3.8]{GLL09}. Indeed, graphs of tree-width at most $k$ can be
represented by binary trees which are called $k$-terms. Moreover, for each
formula $\Psi\in\ICPDL(\Sigma,\Gamma)$ we can construct an $\ICPDL$ formula
$\overline{\Psi}^{k}$ of size $\mathcal{O}(k^{2}|\Psi|)$ over $k$-terms such
that, for all $k$-terms $\tau$, we have $\tau\models\overline{\Psi}^{k}$ iff
$\sem{\tau}\models\Psi$, where $\sem{\tau}$ is the graph denoted by the $k$-term
$\tau$~\cite{mpri-lect}. Hence, satisfiability of $\Psi$ over
$\Graphs^{k}(\Sigma,\Gamma)$ is reduced to satisfiability of
$\overline{\Psi}^{k}$ over $k$-terms.

\subparagraph{Graph interpretation and backward translation~{\rm\cite{CourcelleBook,mpri-lect}}}
The following definitions and results will be used in Section~\ref{sec:DS+time}.
We consider two signatures $(\Sigma,\Gamma)$ and $(\Sigma',\Gamma')$.
Intuitively, a graph $G'\in\Graphs(\Sigma',\Gamma')$ is interpreted in a graph
$G\in\Graphs(\Sigma,\Gamma)$ if we have formulae over the signature
$(\Sigma,\Gamma)$ which, when evaluated on $G$, express nodes, labels and edges
of $G'$. In this paper, we use $\CPDL$ interpretations, which means that the
formulae for the interpretation are in $\CPDL(\Sigma,\Gamma)$. Also, we only
need interpretations when the graphs $G$ and $G'$ have the same set of nodes. In
this simple case, an interpretation $\mathcal{I}$ is given by a tuple of state
formulae $(\sigma_p)_{p\in\Sigma'}$ and a tuple of path formulae
$(\pi_\gamma)_{p\in\Gamma'}$, all in $\CPDL(\Sigma,\Gamma)$. Now, we say that a
graph $G'=(V,E',\lambda')\in\Graphs(\Sigma',\Gamma')$ is
$\mathcal{I}$-interpreted in the graph
$G=(V,E,\lambda)\in\Graphs(\Sigma,\Gamma)$ if, for all $u,v\in V$, all
$p\in\Sigma'$ and all $\gamma\in\Gamma'$, we have
 $p\in\lambda'(u) \text{ iff } G,u\models\sigma_p
 \,\text{ and }\,
 (u,\gamma,v)\in E' \text{ iff } G,u,v\models\pi_\gamma$. 
In this case, we write $G'=\mathcal{I}(G)$.

Interpretations allow for a \emph{backward translation} theorem: for each formula
$\Psi'\in\EQICPDL(\Sigma',\Gamma')$, we can construct a formula
$\Psi\in\EQICPDL(\Sigma,\Gamma)$ such that, for all graphs
$G\in\Graphs(\Sigma,\Gamma)$, we have $\mathcal{I}(G)\models\Psi'$ iff
$G\models\Psi$. The formula $\Psi$ is obtained from $\Psi'$ by replacing the
atomic state formulae $p$ with $\sigma_p$ (for $p\in\Sigma'$) and the atomic
path formulae $\Edge{\gamma}$ with $\pi_\gamma$ (for $\gamma\in\Gamma'$).
Hence, $\Psi$ and $\Psi'$ have same intersection width and
$|\Psi|\leq|\Psi'|\cdot\max\{|\sigma_p|,|\pi_\gamma|\myst
p\in\Sigma',\gamma\in\Gamma'\}$.

\section{Logical definability of realizability}
\label{sec:realizability}
\renewcommand{\TCG}{\Graphs(\emptyset,\Gamma_M)}

\subparagraph{Weighted graphs}
  We consider linear weighted graphs where node labels are irrelevant, i.e., $\Sigma=\emptyset$, and edges are labeled with constraints of the form $<\alpha$ or $\leq \alpha$, where $\alpha\in\mathbb{Z}$, i.e.,
$\Gamma=\{\Succ\}\cup(\{{<},{\leq}\}\times\mathbb{Z})$. Since node labels
are irrelevant, a linear weighted graph is simply denoted $G=(V,E)$.
Often we use a maximal constant $M\in\mathbb{N}$ and let
$\Gamma_M=\{\Succ\}\cup(\{{<},{\leq}\}\times\{-(M-1),\ldots,0,\ldots,M-1\})$. A
graph $G\in\TCG$ is called \emph{$M$ weight-bounded}. If we only compare
using $\leq$, i.e., if there are no edges of the form $(u,<,\alpha,v)$, then we
say that the graph is \emph{closed} or a graph with closed constraints.
Otherwise, we call it a \emph{mixed} weighted graph or a graph with mixed
constraints.

 \begin{figure} [ht] \begin{center}
 \begin{tikzpicture}[->,thick]
 \tikzstyle{dashdotted}=[dash pattern=on 3pt off 2pt on \the\pgflinewidth off 2pt]
 \node[state, draw=white] at (-20,-2) (A1) {$y:=0$} ;
 \node[initial, cir11, initial text ={}] at (-20,0) (A) {} ;

 \node[state, draw=white] at (-18,-2) (B1) {$x:=0$} ;
 \node[cir11] at (-18,0) (B) {};

 \node[state, draw=white] at (-16,-2) (C1) {$\textcolor{green!50!red}{x>2}$} ;
 \node[cir11] at (-16,0) (C) {};

 \node[state, draw=white] at (-14,-2) (D1) {$\begin{array}{c}\textcolor{blue}{3 \leq y-x < 4} \\ x:=0 \end{array}$} ;
 \node[cir11] at (-14,0) (D) {};

 \node[state, draw=white] at (-12,-2) (E1) {$\begin{array}{c}\textcolor{yellow!20!red}{1 < x \leq 3} \\ \textcolor{violet}{y-x \leq 6} \end{array}$} ;
 \node[state, draw=white] at (-12,-0.5) (E2) {} ;
 \node[cir11] at (-12,0) (E) {};

 \path (A) edge node {}node {} (B);
 \path (B) edge node {}node {} (C);
 \path (C) edge node {}node {} (D);
 \path (D) edge node {}node {} (E);

 \draw[dashdotted,-] (-20,-0.5)--(-20,-1.5);
 \draw[dashdotted,-] (-18,-0.5)--(-18,-1.5);
 \draw[dashdotted,-] (-16,-0.5)--(-16,-1.5);
 \draw[dashdotted,-] (-14,-0.5)--(-14,-1.5);
 \draw[dashdotted,-] (-12,-0.5)--(-12,-1.5);

 \path(C) edge[draw=green!50!red,bend left=60] node[below] {$\textcolor{green!50!red}{<-2}$} node{}(B);

 \path(A) edge[draw=blue,bend left=40] node[above] {$\textcolor{blue}{<4}$} node{}(B);
 \path(B) edge[draw=blue,bend left=60] node[below] {$\textcolor{blue}{\leq -3}$} node{}(A);

 \path(A) edge[draw=violet,bend left=60] node[below] {$\textcolor{violet}{\leq 6}$} node{}(D);

 \path(D) edge[draw=yellow!20!red,bend left=60] node[above] {$\textcolor{yellow!20!red}{\leq 3}$} node{}(E);
 \path(E) edge[draw=yellow!20!red,bend left=60] node[below] {$\textcolor{yellow!20!red}{< -1}$} node{}(D);

 \end{tikzpicture}
 \caption{A realizable linear weighted graph obtained from a sequence of instructions
 of a timed system. $x, y$ are real-valued variables called \emph{clocks}.
 $x:=0$ ($y:=0$) denotes reset instructions.
 Changing the last instruction %
 to $y-x
 \leq 5$ gives a non-realizable weighted graph. The non-realizability follows
 from (i) there is a time elapse more than 5 between the first and
 third nodes, (ii) the time elapse is at most 5 between the first and fourth nodes, and
 (iii) time is monotone, hence there is at least zero time elapse
 between the third and fourth nodes.
 This gives a negative cycle between
 the first and fourth nodes.
 }
 \label{fig:wtgph}
\end{center}
\end{figure}
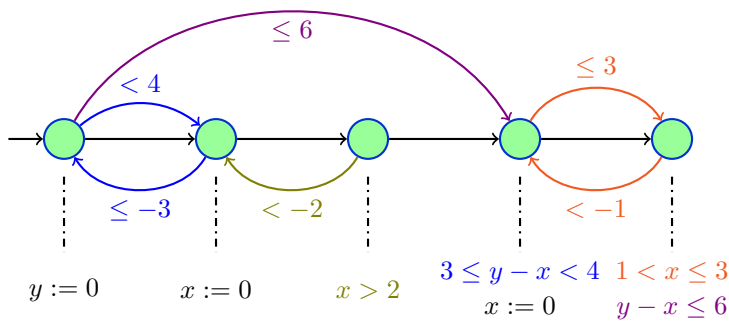 

\subparagraph{Realizability}
One important property of
interest, which is the focus of this paper, is \emph{realizability}. The
property of realizability asks whether the constraints defined by the weights can
be satisfied in a manner that is consistent with the order. 

\begin{definition}
 A weighted graph $G$ is \emph{realizable} if there exists a time-stamp map
 $\ts:V\rightarrow \mathbb{R}$ such that
 (i) all constraints are satisfied:
 $\forall (u,\lleq,\alpha,v)\in E$, ${\ts(v)-\ts(u)}\lleq \alpha$,
 and (ii) $\ts$ is monotone w.r.t. the linear order:
 $\forall u,v\in V$, if $u\po v$, then $\ts(u)\leq\ts(v)$.
\end{definition}

If $G$ is realizable via a map $\ts$, then we say that $\ts$ is a
\emph{realization} of $G$. Note that the monotonicity could have been enforced
by adding more constraint edges: when $u\podot v$ we could have added an edge
$(v,\leq,0,u)$. With these extra constraints, realizability corresponds to
checking the feasibility of the difference constraints. This is a classical
problem on graphs which amounts to checking the absence of a negative cycle
(see~\cite{CLRS} for more details).
There are many algorithms to solve this problem, e.g., the Bellman-Ford shortest
path algorithm.
Finally, as a quick aside, note that if we have reflexive edges
$(u,\lleq,\alpha,u)\in E$, checking realizability for these constraints is
always vacuously true or false for all possible time-stamps, and is easy.
A realizable linear weighted graph obtained from a sequence of instructions
of a timed system is depicted in Figure~\ref{fig:wtgph}.

\subsection{The first main result: logical definability of realizability}
We are interested in properties of (possibly infinite) collections of such graphs, presented in a finite fashion. In particular, we wish to view graphs as being generated by an automaton, i.e., as behaviors of a system, and we wish to reason about this set of graphs. From this automata-theoretic viewpoint, a natural question to ask is whether the properties that we wish to reason about are definable in a certain logic. We focus on the specific property of realizability in weighted graphs and study its definability in \pdl
in our first main result below.
In the next section, we will explain far-reaching consequences of our logical characterization, and in particular its application for checking emptiness of timed systems.
\begin{theorem}\label{thm:linear}
 Realizability is \pdl definable on the set of graphs $\TCG$. The size of the
 formula is polynomial in $M$ and its intersection width is 2.
\end{theorem}

We prove the above theorem in two steps: in Subsection~\ref{sec:closed}, we consider
closed graphs and show that the logical definition is rather easy for them. Then, in Subsection~\ref{sec:mixed}, we consider graphs with mixed constraints.

Throughout the proof, given a linear weighted graph $G=(V,E)$ with $|V|=n$, we
let $V=\{u_1,\ldots,u_n\}$ with $u_1\podot u_2 \podot \cdots \podot u_n$. We start with a simple observation regarding the time-stamps witnessing realizability in weighted graphs. Given an $M$ weight-bounded graph $G=(V,E)$, a mapping $\ts\colon V\rightarrow\mathbb{R}$ is said to be \emph{slowly monotone} if $0 \leq
\intp{\ts(v)}-\intp{\ts(u)}\leq M-1$ whenever $u \podot v$,
where $\intp{x}$ denotes the fractional part of the real number $x$.
If a realization of a graph $G$ is not slowly monotone, then there must exist two consecutive points whose time-stamps are separated by more than $M-1$. But in this case there can be no forward edge (i.e., upper bound) that crosses this point, and hence the time difference between them can be reduced to any value larger than $M-1$ without affecting realizability. Formally,
 \begin{restatable}{lemma}{timestamps}
 \label{lem:time-stamps}
 A graph $G=(V, E)$ in $\TCG$
 is realizable iff there is a slowly monotone map $\ts:V\rightarrow \mathbb{R}$
 that realizes $G$.
 \end{restatable}
\begin{proof}
 Let $G=(V,E)\in\TCG$ be realizable. Then
 there exists a map $\ts'\colon V\to \mathbb{R}$ such that all constraints
 are satisfied and $\ts'$ is monotone w.r.t. $\po$.

 A \emph{large gap} in $\ts'$ is an integer $i < n$ such that
 $\intp{\ts'(u_{i+1})}-\intp{\ts'(u_i)}\geq M$. First, if $\ts'$ has no large
 gap, then $\ts'$ is slowly monotone and we are done. Henceforth, we assume
 that $\ts'$ has at least one large gap, and we prove
 Lemma~\ref{lem:time-stamps} by backward induction on the smallest large gap of
 $\ts'$.

 Let $i$ be the smallest large gap of $\ts'$. Notice that $\ts'(u_{i+1}) -
 \ts'(u_i) > M-1$. Since $\ts'$ is a realization of $G$, there cannot exist a
 forward edge $(u,\lleq,\alpha,v)\in E$ crossing $(u_i,u_{i+1})$, i.e., such
 that $u\po u_i \podot u_{i+1} \po v$, since $\alpha\leq M-1$ contradicts the
 satisfaction of constraints. The back edges $(u,\lleq,\alpha,v)\in E$ crossing
 over $(u_i,u_{i+1})$, i.e., $v\po u_i \podot u_{i+1} \po u$ are all
 satisfied since $\ts'(v)-\ts'(u)\leq\ts'(u_i)-\ts'(u_{i+1})< 1-M \leq \alpha$. Now,
 if we reduce the difference $\ts'(u_{i+1})-\ts'(u_i)$ to any value larger
 than $M-1$, then the constraint on back edges are still satisfied.
 Hence, %
 we choose $t>\ts'(u_i)+M-1$ such that
 $\intp{t}=\intp{\ts'(u_i)}+M-1$ and we define
 \begin{equation*}
 \ts''(w)=\begin{cases}
 \ts'(w) & \text{if } w\po u_i \\
 t + \ts'(w) - \ts'(u_{i+1}) & \text{otherwise.}
 \end{cases}
 \end{equation*}
 We can check that $\ts''$ is a monotone time-stamping satisfying all
 constraints of $G$. Moreover, all large gaps of $\ts''$, if any, are
 greater than $i$. By backward induction,
 we conclude that there exists a monotone time-stamping $\ts$
 satisfying all constraints of $G$ and having no large gap.
 Note indeed that the $\ts$ values themselves can be arbitrarily large, but the
 difference between integral parts of consecutive points is at most $M-1$.
\end{proof}

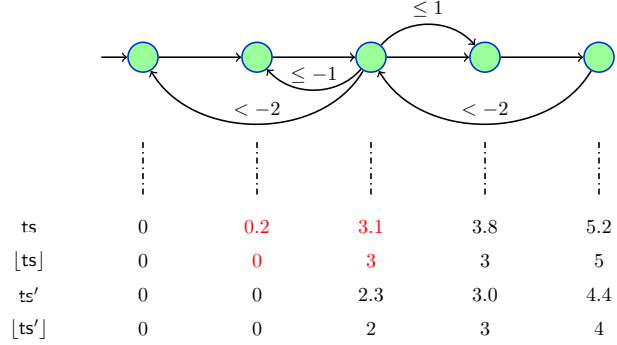
\begin{figure} [ht] \begin{center}
 \scalebox{0.75}{
 \begin{tikzpicture}[->,thick]
 \tikzstyle{dashdotted}=[dash pattern=on 3pt off 2pt on \the\pgflinewidth off 2pt]
 \node[state, draw=white] at (-22,-3) (A1) {$\ts$} ;
 \node[state, draw=white] at (-22,-3.6) (A2) {$\lfloor \ts \rfloor $} ;
 \node[state, draw=white] at (-22,-4.2) (A5) {$\ts'$} ;
 \node[state, draw=white] at (-22,-4.8) (A7) {$\lfloor \ts' \rfloor$} ;

 \node[initial, cir11, initial text ={}] at (-20,0) (A) {} ;
 \node[state, draw=white] at (-20,-3) (A3) {0} ;
 \node[state, draw=white] at (-20,-3.6) (A4) {0} ;
 \node[state, draw=white] at (-20,-4.2) (A6) {0} ;
 \node[state, draw=white] at (-20,-4.8) (A8) {0} ;

 \node[cir11] at (-18,0) (B) {};

 \node[state, draw=white] at (-18,-3) (B3) {\textcolor{red}{0.2}} ;
 \node[state, draw=white] at (-18,-3.6) (B4) {\textcolor{red}{0}} ;
 \node[state, draw=white] at (-18,-4.2) (B5) {0} ;
 \node[state, draw=white] at (-18,-4.8) (B6) {0} ;

 \node[cir11] at (-16,0) (C) {};
 \node[state, draw=white] at (-16,-3) (C3) {\textcolor{red}{3.1}} ;
 \node[state, draw=white] at (-16,-3.6) (C4) {\textcolor{red}{3}} ;
 \node[state, draw=white] at (-16,-4.2) (C5) {2.3} ;
 \node[state, draw=white] at (-16,-4.8) (C6) {2} ;

 \node[cir11] at (-14,0) (D) {};
 \node[state, draw=white] at (-14,-3) (D3) {3.8} ;
 \node[state, draw=white] at (-14,-3.6) (D4) {3} ;
 \node[state, draw=white] at (-14,-4.2) (D5) {3.0} ;
 \node[state, draw=white] at (-14,-4.8) (D6) {3} ;

 \node[cir11] at (-12,0) (E) {};
 \node[state, draw=white] at (-12,-3) (E3) {5.2} ;
 \node[state, draw=white] at (-12,-3.6) (E4) {5} ;
 \node[state, draw=white] at (-12,-4.2) (E5) {4.4} ;
 \node[state, draw=white] at (-12,-4.8) (E6) {4} ;

 \path (A) edge node {}node {} (B);
 \path (B) edge node {}node {} (C);
 \path (C) edge node {}node {} (D);
 \path (D) edge node {}node {} (E);

 \draw[dashdotted,-] (-20,-1.5)--(-20,-2.5);
 \draw[dashdotted,-] (-18,-1.5)--(-18,-2.5);
 \draw[dashdotted,-] (-16,-1.5)--(-16,-2.5);
 \draw[dashdotted,-] (-14,-1.5)--(-14,-2.5);
 \draw[dashdotted,-] (-12,-1.5)--(-12,-2.5);

 \path(C) edge[bend left=60] node[above] {$<-2$} node{}(A);

 \path(C) edge[bend left=50] node[above] {$\leq -1$} node{}(B);

 \path(C) edge[bend left=50] node[above] {$\leq 1$} node{}(D);
 \path(E) edge[bend left=60] node[above] {$< -2$} node{}(C);

 \end{tikzpicture}
 }
 \caption{Replacing large gaps.}
 \label{fig:ts}
\end{center}
\end{figure}

Let us see the above Lemma in action on an example. Consider the weighted graph in Figure~\ref{fig:ts} with map $\ts\colon V\rightarrow \mathbb{R}$ which has a single large gap. We will apply Lemma \ref{lem:time-stamps} to show that we can replace $\ts$ with a slowly monotone
map $\ts'$. For the example, we have $M=3$. The large gap is between time-stamps $0.2$ and $3.1$.
$\alpha$ is chosen as any value $> 0.2+2=2.2$. Let $\alpha=2.3$.
We then replace the time-stamp $3.1$ by $2.3$. Indeed, this new time-stamp satisfies
the constraint $\leq -1$ of the gap ($2.3-0.2 \geq 1$).
Had we tried any other time-stamp satisfying the constraint $\leq -1$,
for instance $1.3$ instead of $2.3$,
we might fail to satisfy the constraint $<-2$ between the first and third time-stamps.
Thus, reducing the time difference to be just more than $M-1$ is a safe choice
whenever we have a large gap. We propagate the reduction by $0.8$ ($3.1 \mapsto 2.3$)
to the subsequent time-stamps as well, so that the relative time differences
are not affected. This gives us the slowly monotone map $\ts'$.

Next, we have a crucial definition on general weighted graphs. Given an $M$
weight-bounded linear graph $G=(V,E)$, a
\emph{time-stamping modulo $M$} is a map $\tsm\colon V\rightarrow
\mathbb{Z}_M=\{0,\ldots M-1\}$. For all $u,v\in V$, we set $\dtsm(u,v)=
(\tsm(v)-\tsm(u)) \mod M$. Further, $(u,v)$ is said to be \emph{$\tsm$-big} if
there exist $w_1,w_2\in V$ such that $u\po w_1\prec w_2\po v$ and
$\dtsm(u,w_1)+\dtsm(w_1,w_2)\geq M$. Observe that, if $v \po u$, then
$(u,v)$ cannot be $\tsm$-big.

\begin{definition}\label{def:weak-realizability}
 A time-stamping modulo $M$ $\tsm$ is said to \emph{weakly satisfy}
 $G=(V,E)$ if for all $e=(u,\lleq,\alpha,v)\in E$,
 \begin{itemize}[nosep]
 \item[(a)] if $u\po v$, then $(u,v)$ is not $\tsm$-big and $\dtsm(u,v)\leq \alpha$;
 \item[(b)] if $v\prec u$ then $(v,u)$ is $\tsm$-big or $\dtsm(v,u)\geq -\alpha$.
 \end{itemize}
\end{definition}

Lemma~\ref{lem:weak-realizability1}
below shows that for linear weighted graphs, existence of such a
map is a necessary condition for realizability.
But first, we establish some useful facts.
Recall that $V=\{u_1,\ldots,u_n\}$ with $u_1\podot u_2 \podot \cdots \podot u_n$.
For $i\leq j$, we also define
$\dtsmp(u_i,u_j)=\min\{M,\dtsm(u_i,u_{i+1})+\cdots+\dtsm(u_{j-1},u_j)\}$ and
$\dtsmp(u_j,u_i)= -\dtsmp(u_i,u_j)$. Notice that we have $\dtsmp(u_i,u_i)=0$.

\begin{restatable}{claim}{useful}
 \label{claim:useful}
 Let $G=(V,E) \in \TCG$ %
 and let $\ts\colon V\to \mathbb{R}$ be a slowly monotone map (which need not satisfy the constraints of $G$). Define $\tsm\colon V\to
 \mathbb{Z}_M$ by $\tsm(v)=\intp{\ts(v)}\mod M$ for all $v\in V$. Then, for
 all $u,v\in V$ such that $u \po v$, we have $\dtsmp(u,v) =
 \min\{\intp{\ts(v)}-\intp{\ts(u)},M\}$. Furthermore, we have $\dtsmp(u,v) =
 M$ if $(u,v)$ is $\tsm$-big, and $\dtsmp(u,v) = \dtsm(u,v)$ otherwise.
\end{restatable}

\begin{proof}
 First, $\dtsm(u_i,u_j)=(\intp{\ts(u_j)}-\intp{\ts(u_i)}\mod M)\leq
 \intp{\ts(u_j)}-\intp{\ts(u_i)}$ for all $i\leq j$.
 Since $\ts$ is slowly monotone, we know that
 $\intp{\ts(u_{i+1})}-\intp{\ts(u_i)} < M$ for all $i < n$. We deduce that
 $\dtsm(u_i,u_{i+1}) = \intp{\ts(u_{i+1})}-\intp{\ts(u_i)}$. Hence, from the
 definition of $\dtsmp$, we have $\dtsmp(u_i,u_j) =
 \min\{M,\intp{\ts(u_j)}-\intp{\ts(u_i)}\}$ for $i\leq j$, and $\dtsm(u_i,u_j)
 = \dtsmp(u_i,u_j)$ if and only if $\dtsmp(u_i,u_j) < M$.
 Moreover, if $(u_i,u_j)$ is $\tsm$-big, there exist integers $k$ and $\ell$ such that
 $i \leq k < \ell \leq j$ and $\dtsm(u_i,u_k) + \dtsm(u_k,u_\ell) \geq M$.
 Hence $\intp{\ts(u_j)} \geq \intp{\ts(u_\ell)} \geq \intp{\ts(u_k)} + \dtsm(u_k,u_\ell) \geq
 \intp{\ts(u_i)} + \dtsm(u_i,u_\ell) + \dtsm(u_k,u_\ell) \geq \intp{\ts(u_i)} + M$,
 and thus we have $\dtsmp(u_i,u_j) = M$.

 Conversely, if $\dtsmp(u_i,u_j) = M$, then $\intp{\ts(u_j)} \geq \intp{\ts(u_i)} + M$.
 Then, let $k$ be the smallest integer such that $k \geq i$ and $\intp{\ts(u_k)} \geq \intp{\ts(u_i)} + M$.
 We must have $k > i$.
 It follows that $\intp{\ts(u_{k-1})} < \intp{\ts(u_i)} + M$, and therefore that
 $\intp{\ts(u_{k-1})} - \intp{\ts(u_i)} = \dtsm(u_i,u_{k-1})$.
 Since $\ts$ is slowly monotone, we also have $\intp{\ts(u_k)} < \intp{\ts(u_{k-1})} + M$,
 and therefore $\intp{\ts(u_k)} - \intp{\ts(u_{k-1})} = \dtsm(u_{k-1},u_k)$.
 This shows that $\dtsm(u_i,u_{k-1}) + \dtsm(u_{k-1},u_k) = \intp{\ts(u_k)} - \intp{\ts(u_i)} \geq M$,
 and thus that $(u_i,u_j)$ is $\tsm$-big, which completes the proof.
\end{proof}

Given that $|\alpha|<M$ for all edges $e = (u,\lleq,\alpha,v) \in E$,
Claim~\ref{claim:useful} provides us with the following, alternative
characterization of weak satisfiability. 

\begin{restatable}{lemma}{weakrealizability}
 \label{cor:weak-realizability}
 A time-stamping modulo $M$ $\tsm$ weakly satisfies the graph $G=(V,E)$ if and
 only if $\dtsmp(u,v) \leq \alpha$ for all $(u,\lleq,\alpha,v)\in E$.
\end{restatable}

\begin{proof}
 Let $(u,\lleq,\alpha,v)\in E$ with $u\po v$. If $(u,v)$ is not
 $\dtsm$-big and $\dtsm(u,v)\leq\alpha$, then
 $\dtsmp(u,v)=\dtsm(u,v)\leq\alpha$. Conversely, if $\dtsmp(u,v)\leq\alpha$,
 since $\alpha<M$ we deduce that $(u,v)$ is not $\dtsm$-big and
 $\dtsm(u,v)=\dtsmp(u,v)\leq\alpha$.

 Then, let $(u,\lleq,\alpha,v)\in E$ with $v\prec u$. If $(v,u)$ is $\dtsm$-big,
 then $\dtsmp(u,v)=-\dtsmp(v,u)=-M<\alpha$.
 Likewise, if $(v,u)$ is not $\dtsm$-big and
 $\dtsm(v,u)\geq-\alpha$, then
 $\dtsmp(u,v)=-\dtsmp(v,u)=-\dtsm(v,u)\leq\alpha$. Conversely, if
 $\dtsmp(u,v)\leq\alpha$, then either $\dtsmp(u,v)=-M$ and $(v,u)$ is
 $\dtsm$-big, or else $(v,u)$ is not $\dtsm$-big and
 $\dtsm(v,u)=\dtsmp(v,u)=-\dtsmp(u,v)\geq-\alpha$.
\end{proof}

Now, we obtain one direction of the characterization, which works both for closed and open constraints.
\begin{lemma}\label{lem:weak-realizability1}
 If $G \in \TCG$ %
 is realizable, then
 there exists a time-stamping modulo $M$ that weakly satisfies $G$.
\end{lemma}

\begin{proof}
 Lemma~\ref{lem:time-stamps} proves that there exists a slowly monotone
 time-stamping $\ts$ that satisfies the constraints $G$. We define $\tsm\colon
 V\rightarrow \mathbb{Z}_M$ by $\tsm(v)=\intp{\ts(v)}\mod M$, and we show below
 that $\tsm$ weakly satisfies $G$.

 Let $(u,\lleq,\alpha,v)\in E$. By
 Lemma~\ref{cor:weak-realizability}, it is enough to show that
 $\dtsmp(u,v)\leq\alpha$. According to Claim~\ref{claim:useful},
 distinguishing the cases $u\po v$ and $v\prec u$, we show easily that
 $\dtsmp(u,v) \leq \intp{\ts(v)}-\intp{\ts(u)}$ or $\dtsmp(u,v) = -M$. In the
 first case, it follows that $\dtsmp(u,v) \leq \intp{\ts(v)}-\intp{\ts(u)} \leq
 \ts(v) - \intp{\ts(u)} < \ts(v) - \ts(u) + 1 \leq \alpha + 1$, and in the
 second case, we also have $\dtsmp(u,v) = -M < \alpha + 1$. Hence, in both
 cases, we have $\dtsmp(u,v) < \alpha + 1$. Observing that $\dtsmp(u,v)$ and
 $\alpha$ are integers proves that $\dtsmp(u,v) \leq \alpha$.
\end{proof}

The converse of the above lemma does not hold with mixed guards and this will be
handled in the next subsection. However, for closed guards it yields the following characterization.

\subsubsection{Characterizing realizability in closed graphs}\label{sec:closed}

\begin{lemma}\label{lem:weak-realizability2}
 A closed graph $G=(V,E)$ in $\TCG$
 is realizable iff there exists a time-stamping modulo $M$ that weakly
 satisfies $G$.
\end{lemma}

\begin{proof}
 One direction is Lemma~\ref{lem:weak-realizability1}. Conversely, suppose that
 $\tsm\colon V \rightarrow \mathbb{Z}_M$ is a time-stamping modulo $M$ that weakly
 satisfies $G$. Then, the map $\ts\colon V \rightarrow \mathbb{N}$ defined
 inductively by $\ts(u_1) = 0$ and $\ts(u_{i+1}) = \ts(u_i) + \dtsm(u_i,u_{i+1})$
 is a slowly monotone map.

 Let $(u,\lleq,\alpha,v) \in E$. By Claim~\ref{claim:useful}, distinguishing the
 cases $u\po v$ and $v\prec u$, we show easily that $\dtsmp(u,v) \geq
 \intp{\ts(v)}-\intp{\ts(u)}$ or $\dtsmp(u,v)=M$. Since $\tsm$ weakly
 satisfies $G$ (i.e., $\dtsmp(u,v) \leq \alpha$) and $M > \alpha$, the second
 case is impossible. It follows that $\ts(v) - \ts(u) = \intp{\ts(v)} -
 \intp{\ts(u)} \leq \dtsmp(u,v) \leq \alpha$, which shows that $\ts$ satisfies
 the constraints of $G$.
\end{proof}

It remains to encode the characterization of Lemma~\ref{lem:weak-realizability2}
in \pdl to obtain the logical definability of realizability for linear weighted
graphs.

\subparagraph{$\EQLCPDL$ characterization}

We use existential quantification over atomic propositions
$p_0,\ldots,p_{M-1}$ to guess the time-stamping modulo $M$. Intuitively, a node
satisfies $p_i$ iff its $\tsm$ value is $i$. So we define the formula
$
\exists p_0,\ldots,p_{M-1}~
\mathsf{Partition}\wedge\mathsf{Forward}\wedge\mathsf{Backward}
$ 
where the auxiliary formulae are defined in Figure~\ref{tab:pdl-closed}. The
formula $\mathsf{Partition}$ states that every vertex satisfies exactly one
$p_i$ ($0\leq i<M$).

For $0\leq i,j<M$, let $\delta_M(i,j)=(j-i)\mod M$. We use a path formula to
characterize pairs of vertices that are $\tsm$-big: a pair $(u,v)$ is
$\tsm$-big iff we can go from node $u$ to node $v$ following the path formula
$\mathsf{BigPath}$. 

Since negation is not allowed at the level of path formulae, we provide another
formula, $\mathsf{SmallPath}$, to express that a pair $(u,v)$ of vertices is not
$\tsm$-big. There are two cases, depending on whether $\tsm(u) \leq \tsm(v)$ or
not. In both cases, $(u,v) \models \mathsf{SmallPath}_{i,j}$ iff $u \po v$,
$(u,v)$ is not $\tsm$-big, $i=\tsm(u)$ and $j=\tsm(v)$.

Formulae $\mathsf{Forward}$ and $\mathsf{Backward}$ respectively state the two
conditions in Definition~\ref{def:weak-realizability}. The constraint on
$\po$-forward edges is stated using the loop operator of $\LCPDL$. By excluding
the existence of a loop following the path
$\mathsf{BigPath}\cdot{\Edgem{\leq\alpha}}$ we make sure that forward edges
$(u,v)\in E_{\leq\alpha}$ are not $\tsm$-big.
Now, to ensure that forward edges $(u,\lleq,\alpha,v)$ satisfy
$\dtsm(u,v)\leq\alpha$, we exclude the existence of a path violating this
property, i.e., a loop following $\test{p_i} \cdot {\Edge{\leq\alpha}} \cdot \test{p_j}
\cdot ({\proc}^{-1})^+$ with $\delta_M(i,j)>\alpha$.

\begin{figure}
\begin{align*}
 \mathsf{Partition} &= \forallnodes \bigvee_{0\leq i<M} [p_i \wedge
 \bigwedge_{j\neq i} \neg p_j]
 \\
 \mathsf{BigPath} &=
 \hspace{-8mm}\sum_{\substack{0\leq i,j,k<M \\ \delta_M(i,j)+\delta_M(j,k)\geq M}}\hspace{-8mm}
 \test{p_i} \cdot {\proc}^+ \cdot \test{p_j} \cdot {\proc}^+ \cdot \test{p_k} \cdot {\proc}^\ast
 \\
 \mathsf{SmallPath}_{i,j} &= \test{p_i} \cdot \Big(\hspace{-3mm}\sum_{i\leq k \leq \ell \leq j}
 \hspace{-4mm}
 \test{p_k} \cdot {\proc} \cdot \test{p_\ell}\Big)^\ast \hspace{-1mm}\cdot \test{p_j}
 ~\text{if}~ i\leq j
 \\
 \mathsf{SmallPath}_{i,j} &= \hspace{-6mm}\sum_{0 \leq \ell \leq j < i \leq k < M}\hspace{-6mm}
 \mathsf{SmallPath}_{i,k} \cdot {\proc} \cdot \mathsf{SmallPath}_{\ell,j}
 \qquad\quad~\text{if } j<i \,
\\
 \mathsf{Forward} &= \neg\existsnode \hspace{-2mm}\bigvee_{-M<\alpha<M}\hspace{-2mm}
 \existsloop{\mathsf{BigPath}\cdot{\Edgem{\leq\alpha}}}
 \\
 &{}\wedge\neg\existsnode\hspace{-2mm}\bigvee_{\substack{0\leq i,j<M \\ \delta_M(i,j)>\alpha}}\hspace{-2mm}
 \existsloop{\test{p_i} \cdot {\Edge{\leq\alpha}} \cdot \test{p_j} \cdot ({\proc}^{-1})^+}
 \\
 \mathsf{Backward} &= \neg\existsnode
 \hspace{-3mm}\bigvee_{\substack{-M<\alpha<M \\ 0\leq i,j<M \\ \delta_M(i,j)<-\alpha}}\hspace{-2mm}
 \existsloop{ \mathsf{SmallPath}_{i,j} \cdot {\Edge{\leq\alpha}}}
 \end{align*}
\caption{$\LCPDL$ for realizability of linear closed graphs}
\protect\label{tab:pdl-closed}
\end{figure}

\subsubsection{A characterization with mixed guards}\label{sec:mixed}
The characterization above is not sufficient when some of the constraints are
strict, i.e., $E$ contains edges of the form $(u,<,\alpha,v)$. It turns out that we
need an additional condition to make sure that the fractional parts do not
violate the realizability.

\begin{definition}\label{dfn:lef-ltf}
 Given a graph $G=(V,E)$ and a time-stamping $\tsm:V\rightarrow \mathbb{Z}_M$ modulo $M$,
 we define two binary relations $\lef$ and $\ltf$ on $V$:
 \begin{itemize}[nosep]
 \item $(u,v)\in \lef$ iff one of the following conditions hold:
 \begin{enumerate}
 \item\label{item:c3} $u\pos v$, $(u,v)$ is not $\tsm$-big and
 $\dtsm(u,v)=\alpha$ for some edge $(u,\lleq,\alpha,v)\in E$;
 \item\label{item:c2} $v\pos u$, $(v,u)$ is not $\tsm$-big and
 $\dtsm(v,u)=-\alpha$ for some edge $(u,\lleq,\alpha,v)\in E$;
 \item\label{item:c1} $v\podot u$ and $\dtsm(u,v)=0$.
 \end{enumerate}
 \item $(u,v)\in \ltf$ iff one of the following conditions hold:
 \begin{enumerate}
 \item\label{item:c3s} $u\pos v$, $(u,v)$ is not $\tsm$-big and
 $\dtsm(u,v)=\alpha$ for some edge $(u,{<},\alpha,v)\in E$;
 \item\label{item:c2s} $v\pos u$, $(v,u)$ is not $\tsm$-big and
 $\dtsm(v,u)=-\alpha$ for some edge $(u,{<},\alpha,v)\in E$.
 \end{enumerate}
 \end{itemize}
\end{definition}

Notice that $\ltf \subseteq \lef$.
The idea is that these relations give the ordering between the fractional parts.
Thus, $(u,v)\in\lef$ (resp. $\ltf$) means that the fractional part of $\ts(u)$
must be at least (resp. strictly greater than) the fractional part of $\ts(v)$.
Once again, since $|\alpha| < M$ for all edges
$(u,\lleq,\alpha,v) \in
E$, Claim~\ref{claim:useful} provides an alternative characterization of the
relations $\lef$ and $\ltf$.

\begin{lemma}
 Consider graph $G=(V,E)$, $\tsm:V\rightarrow \mathbb{Z}_M$
 modulo $M$ and a pair $(u,v)$ of vertices of $G$. Then,
 \begin{itemize}
 \item $(u,v) \in \lef$ iff there exists an edge $(u,\lleq,\alpha,v)\in E$ such that
 $\dtsmp(u,v) = \alpha$, or $v \podot u$ and $\dtsmp(u,v) = 0$;
 \item $(u,v) \in \ltf$ iff there exists an edge $(u,{<},\alpha,v)\in E$ such that $\dtsmp(u,v) = \alpha$.
 \end{itemize}
\end{lemma}

\begin{lemma}\label{lem:open-charac}
 Let $G=(V,E)$ be an $M$ weight-bounded graph with a linear order and mixed constraints. $G$ is realizable iff there exists a time-stamping modulo $M$ $\tsm$ such that
 (i) $\tsm$ weakly satisfies $G$ \emph{and} (ii) there do not exist $u,v\in V$ such that $(u,v)\in\ltf$ and $(v,u)\in \lef^\ast$, where $\lef^\ast$ is the reflexive transitive closure of $\lef$.
\end{lemma}

\begin{proof}
In the forward direction, let $G$ be realizable. Let $\ts\colon V \rightarrow \mathbb{R}$
be a slowly monotone map that realizes $G$, and let $\tsm$ be the time-stamping modulo $M$
defined by $\tsm\colon v \rightarrow \intp{\ts(v)} \mod M$.
Lemma~\ref{lem:weak-realizability1} proves that $\tsm$ weakly realizes $G$.
We further claim that, if $(u, v) \in \lef$, then $\fracp{\ts(u)} \geq \fracp{\ts(v)}$,
and that, if $(u, v) \in \ltf$, then $\fracp{\ts(u)} > \fracp{\ts(v)}$.
The proof is as follows.

\begin{itemize}
\item If $(u, v) \in \lef$ because $v \podot u$ and $\dtsmp(u,v) = 0$, then
$\ts(v) \leq \ts(u)$ and $0 = \dtsmp(u,v) = \max\{\intp{\ts(v)} - \intp{\ts(u)},-M\}$.
Hence, $\intp{\ts(u)} = \intp{\ts(v)}$, and therefore $\fracp{\ts(v)} \leq \fracp{\ts(u)}$.

\item If $(u, v) \in \lef$ because there exists an edge $(u,\lleq,$ $\alpha,v) \in E$ such that
$\dtsmp(u,v) = \alpha$, then $-M < \alpha = \dtsmp(u,v) < M$, and
Claim~\ref{claim:useful} proves that $\alpha = \dtsmp(u,v) = \intp{\ts(v)} - \intp{\ts(u)}$.
It follows that $\fracp{\ts(v)} = \ts(v) - \intp{\ts(v)} \leq \ts(u) + \alpha - \intp{\ts(v)} = \ts(u) - \intp{\ts(u)} = \fracp{\ts(u)}$.

\item If $(u, v) \in \ltf$, then the same argument proves that
$\alpha = \dtsmp(u,v) = \intp{\ts(v)} - \intp{\ts(u)}$, and
it follows that $\fracp{\ts(v)} = \ts(v) - \intp{\ts(v)} < \ts(u) + \alpha - \intp{\ts(v)} = \ts(u) - \intp{\ts(u)} = \fracp{\ts(u)}$.
\end{itemize}

In the reverse direction, let $\tsm\colon V \rightarrow \mathbb{Z}_M$ be a time-stamping modulo $M$
that weakly satisfies $G$ and such that (ii) holds.
As a direct consequence of (ii), every path in the graph $G_{\lef} = (V, \lef)$
contains at most $|V|$ edges in $\ltf$. Indeed, otherwise two such edges would start from the same vertex,
so that one edge would belong to a cycle of $G_{\lef}$. Hence, for every vertex $v \in V$, we define
the integer $\ts_1(v)$ as the largest number of edges in $\ltf$ that may be used
by a path in $G_{\lef}$ starting from $v$: observe that $0 \leq \ts_1(v) \leq |V|$.

By construction, for every pair $(u,v)$ in $\lef$,
we have $\ts_1(u) \geq \ts_1(v)$, and we even have $\ts_1(u) > \ts_1(v)$ if $(u, v) \in \ltf$.
Then, consider the map $\ts_0\colon V \rightarrow \mathbb{N}$ defined inductively by $\ts_0(u_1) = 0$
and $\ts_0(u_{i+1}) = \ts_0(u_i) + \dtsm(u_i,u_{i+1})$. The proof of Lemma~\ref{lem:weak-realizability2} shows
that $\ts_0$ is a slowly monotone map and that $\ts_0(v) - \ts_0(u) \leq \alpha$ for all edges
$(u,\lleq,\alpha,v) \in E$.

We prove now that the map $\ts\colon V \rightarrow \mathbb{R}$ defined by $\ts(v) = \ts_0(v) + \ts_1(v) / (|V|+1)$
is monotone. For all pairs $(u,v)$,
\begin{itemize}
 \item if $u \podot v$ and $(v,u) \in \lef$, then $\ts(v) = \ts_0(v) + \ts_1(v) / (|V|+1) \geq \ts_0(u) + \ts_1(u) / (|V|+1) \geq \ts(u)$,
 because $\ts_0(v) \geq \ts_0(u)$ and $\ts_1(v) \geq \ts_1(u)$;
 \item if $u \podot v$ and $(v,u) \notin \lef$, then $\dtsmp(v,u) \neq 0$, and therefore $\dtsmp(u,v) \geq 1$, which proves that
 $\ts(v) \geq \ts_0(v) = \ts_0(u) + \dtsmp(u) \geq \ts_0(u) + 1 > \ts_0(u) + \ts_1(u) / (|V|+1) = \ts(u)$.
\end{itemize}

Then, we prove that $\ts$ satisfies the constraints of $G$. Indeed, for every edge
$(u,\lleq,\alpha,v) \in E$,
\begin{itemize}
 \item if $\dtsmp(u,v) = \alpha$, then $(u,v) \in \lef$, and therefore $\ts_1(v) \leq \ts_1(u)$; it follows that
 $\ts(v) = \ts_0(v) + \ts_1(v) / (|V|+1) \leq \ts_0(u) + \alpha + \ts_1(u) / (|V|+1) = \ts(u) + \alpha$;
 \item if $\dtsmp(u,v) = \alpha$ and, furthermore, ${\lleq}={<}$, then $(u,v) \in \ltf$, hence $\ts_1(v) < \ts_1(u)$;
 it follows that
 $\ts(v) = \ts_0(v) + \ts_1(v) / (|V|+1) < \ts_0(u) + \alpha + \ts_1(u) / (|V|+1) = \ts(u) + \alpha$;
 \item if $\dtsmp(u,v) \neq \alpha$, then $\dtsmp(u,v) \leq \alpha-1$, since $\tsm$ weakly satisfies $G$; it follows that
 $\ts(v) = \ts_0(v) + \ts_1(v) / (|V|+1) < \ts_0(v) + 1 \leq \ts_0(u) + (\alpha-1) + 1 \leq \ts(u) + \alpha$.
\end{itemize}
Consequently, in all cases, we have $\ts(v) - \ts(u) \lleq \alpha$,
which completes the proof.
\end{proof}

\subparagraph{\pdl characterization}

As before, we use existentially quantified propositional variables
$p_0,\ldots,p_{M-1}$ to guess the $\tsm$ values. To state weak-realizability,
we use the formula $\mathsf{WRealizable}=
\mathsf{Partition} \wedge \mathsf{Forward} \wedge \mathsf{Backward}$ where the
subformulae have been defined in Figure~\ref{tab:pdl-closed}.
In addition, we have to check the absence of a cycle among the fractional parts,
which contains at least one strict inequality and other, possibly non-strict,
inequalities. By Lemma~\ref{lem:open-charac}, this suffices to ensure
realizability. To capture the ordering among the fractional parts, we use two
\pdl formulae, $\ltf$ and $\lef$ respectively for the strict and non-strict
parts, formally defined in Figure~\ref{tab:pdl-open}.
The \pdl formula $\mathsf{Realizable}$ is then:
\begin{align*}
\exists p_0,\ldots p_{M-1} ~\mathsf{WRealizable}\wedge \neg
\existsnode \existsloop{\ltf\cdot\lef^\ast}
\label{eq:pdl-seq-open}
\end{align*}
The intersection width of $\ltf$ and $\lef$ is 2. Hence, the intersection
width of $\mathsf{Realizable}$ is also 2.
This completes the proof of Theorem~\ref{thm:linear}.

\begin{figure}
 \begin{align*}
 \ford
 &= (\Edge{\leq \alpha}+\Edge{< \alpha}) \cap
 \Big(\hspace{-1mm}\sum_{\substack{0\leq i,j<M \\ \delta_M(i,j)=\alpha}}\hspace{-3mm}
 \mathsf{SmallPath}_{i,j}
 +\hspace{-4mm}\sum_{\substack{0\leq i,j<M \\ \delta_M(j,i)=-\alpha}}\hspace{-3mm}
 \mathsf{SmallPath}_{j,i}^{-1} \Big)
 \\
 &\hspace{5mm} + \hspace{0mm}
 \sum_{i<M} \test{p_i} \cdot {\proc}^{-1} \cdot \test{p_i}
 \\
 \fords &= {\Edge{< \alpha}} \cap \Big(\hspace{-1mm}
 \sum_{\substack{0\leq i,j<M \\ \delta_M(i,j)=\alpha}}
 \hspace{-3mm} \mathsf{SmallPath}_{i,j} +
 \hspace{-4mm}\sum_{\substack{0\leq i,j<M \\ \delta_M(j,i)=-\alpha}}
 \hspace{-3mm} \mathsf{SmallPath}_{j,i}^{-1} \Big)
 \end{align*}
 \caption{$\ICPDL$ formulae for capturing strict guards}
 \protect\label{tab:pdl-open}
\end{figure}

\subsection{Realizability is beyond logical definability in general}
\label{sec-neg}
Above, we have seen the $\pdl$ definability of realizability for linear weighted graphs. In the absence of a linear order, we now show that this is no longer true, even if one uses the strictly more expressive \mso logic (an easy example is the property of connectivity which separates \pdl from \mso). 

We start by defining MSO interpretations, which will be used to formalize the arguments below.
\begin{definition}
\label{def:msoi}
 An MSO interpretation~\cite{CourcelleBook} is a partial function that
 constructs for a given family of input structures, a new family of output
 structures as specified by a number of MSO formulae. The universe of the
 output structure is determined in terms of the universe of the input structure
 as specified by some MSO formula. Each predicate $R(x_1, \dots, x_k)$ in the
 output structure is determined using an MSO formula $\psi^R(x_1, \dots, x_k)$
 over the input structure.
 More precisely, a deterministic MSO interpretation $\tau\colon \Ss \rightarrow \Tt$ is given by
 (i) an MSO sentence $\varphi_{\dom}$ which determines which input structures 
 $S\in\Ss$ are in the domain of $\tau$,
 (ii) an MSO formula $\varphi(x)$ over the signature of $S$, with one free
 variable $x$, which determines the universe of $\tau(S)$ for each $S\in\dom(\tau)$,
 (iii) for each predicate $R$ of arity $k$ of the output signature, a formula
 $\psi^R(x_1, \dots, x_k)$ over the signature of $S$, with $k$ free first order
 variables $x_1, \dots, x_k$, which determines $R$ in $\tau(S)$ as the set of 
 tuples $(x_1,\ldots,x_k)$ from the universe of $\tau(S)$ which satisfy 
 $\psi^R$.
 \end{definition}
For ease of understanding, we give
here an example that illustrates MSO interpretations.

\begin{example}
 Consider as input the family of word structures over alphabet $\{a,b\}$ and
 binary relation $S$ (successor) that satisfy the formula
 $\varphi_{\dom}=\mathsf{is}\_\mathsf{word} \wedge \psi$, where
 $\mathsf{is}\_\mathsf{word}$
 is an \mso sentence stating that $S$ is the successor relation of a total
 order and that each vertex is labeled either $a$ ($\lab_a$) or $b$ ($\lab_b$).
 The formula $\psi$ is given by 
 $$
 \exists x.[\mathsf{first}(x) \wedge \lab_a(x)] \wedge
 \exists x.[\mathsf{last}(x) \wedge \lab_b(x)] \wedge
 \forall x \forall y [\lab_b(x) \wedge S(x,y) \Rightarrow \lab_b(y))]
 $$
 where $\mathsf{first}(x)=\neg \exists z S(z,x)$ and $\mathsf{last}(x)=\neg
 \exists z S(x,z)$.

 Clearly, the input consists of words in $a^+b^+$. Consider
 the MSO interpretation having formulae $\varphi(u)=true$, which asserts that
 the universe is unchanged, formulae $\psi^{\lab_a}(u)=\lab_a(u)$ and
 $\psi^{\lab_b}(u)=\lab_b(u)$, which preserve the labeling of positions as they
 were in the input word, and formulae $\psi^S(u,v)=S(v,u)$, which revert the
 successor edges. Thus, an input word $a^kb^j$ will result in $b^ja^k$ after
 interpretation.
\label{eg-msot}
\end{example}

Next, we recall the backwards translation theorem~\cite{CourcelleBook}, which is used in the proof of Theorem~\ref{thm:non-mso}.
\begin{theorem}[Backwards Translation Theorem,~\cite{CourcelleBook}]
Let $L \subseteq \Gg_2$ be definable in \mso and let $\theta\colon \Gg_1 \rightarrow \Gg_2$ be an \mso interpretation.	
Then the set $\theta^{-1}(L)=\{G \in \Gg_1 \tq \theta(G) \cap L \neq \emptyset\}$ is definable in \mso.
\label{thm:btt}
\end{theorem}

\begin{figure}[ht]
 \begin{center}
 \begin{tabular}{|c|c|}
 \hline
 & \\[-3mm]
 \raisebox{6.5mm}{\includegraphics[scale=1,page=3]{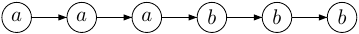}} &
 \includegraphics[scale=1,page=5]{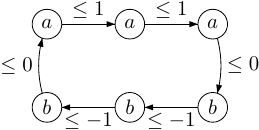} \\
 \hline
 & \\[-3mm]
 \raisebox{6.5mm}{\includegraphics[scale=1,page=4]{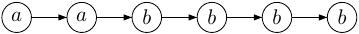}} &
 \includegraphics[scale=1,page=6]{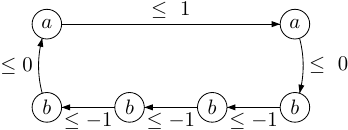} \\
 \hline
 \end{tabular}
 \end{center}
 \caption{The MSO interpretation that interprets words $a^n b^m$ as realizable
 weighted graphs iff $n \geq m$.}
 \label{fig:two-lin}
\end{figure}

 \begin{restatable}{theorem}{nonmso}
 The property of realizability is not definable in \mso for weighted graphs
 without the linear order.
 \label{thm:non-mso}
 \end{restatable}
\begin{proof}
 We prove the result by contradiction using MSO interpretations. Let us
 consider word structures defined by the formula $\varphi_\dom{}$ in Example
 ~\ref{eg-msot}. We define the MSO interpretation $\theta$ that takes as input
 the above family of word structures. We construct a family of weighted graph
 structures as our output. The following MSO formulae complete the
 interpretation. The predicates $\edg{\bow \beta}$ in the output structure
 are determined by formulae $\psi^{\edg{\bow \beta}}$ over the signature
 of the input word structure.
 \begin{enumerate}[nosep]
 \item $\varphi(u)=true$. This ensures that all nodes of the input word
 are also part of the output graph.

 \item $\psi^{\edg{\leq 1}}(u,v)= S(u,v) \wedge \lab_a(u) \wedge \lab_a(v)$

 \item $\psi^{\edg{\leq -1}}(u,v)= S(u,v) \wedge \lab_b(u) \wedge \lab_b(v)$

 \item $\psi^{\edg{ \leq 0}}(u,v)= [S(u,v) \wedge \lab_a(u) \wedge \lab_b(v)]
 \vee [\mathsf{last}(u) \wedge \mathsf{first}(v)]$,

 \item $\varphi^{\podot}(u,v)=[S(u,v) \wedge \lab_a(u) \wedge \lab_a(v)]
 \vee [S(v,u) \wedge \lab_b(u) \wedge \lab_b(v)]$
 \end{enumerate}
Figure~\ref{fig:two-lin} illustrates this interpretation by giving two input
words and the respective weighted graphs obtained. It can be seen that if one
starts from words of the form $a^n b^m$ where $n \geq m$, then the resulting
graph is realizable, and otherwise, it is not since there is a negative cycle.
If we consider $L \subseteq \Gg$ to be the set of realizable graphs, and assume
that $L$ is definable in \mso, then by the Backwards translation theorem, we obtain
$\theta^{-1}(L)=\{a^n b^m \tq n \geq m\}$ to be a language definable in \mso, which
we know is not the case. Hence, realizability is not an \mso-definable property
of weighted graphs. Notice that, by the formula $\varphi^{\podot}(u,v)$, the
weighted graphs constructed are not linear but are covered by two chains.
\end{proof}

\section{Analyzing timed systems with data structures}
\label{sec:conseq}
In this section, we develop a generic technique to analyze timed systems with auxiliary data structures. We start with untimed systems with data structures.

\subsection{Capturing data structure operations as graphs}

\begin{figure}[th!]
 \scalebox{0.58}{
 \begin{tikzpicture}[->,thick]
 \node[initial, cir, initial text ={}] at (11,0) (aA) {$\{\nop\}$} ;

 \node[cir] at (13,0) (aB) {$\{w(d_1)\}$};

 \node[cir] at (15,0) (aC) {$\{w(d_2)\}$};

 \node[cir] at (17,0) (aD) {$\{w(d_2)\}$};

 \node[cir] at (19,0) (aE) {$\{r(d_1)\}$};

 \node[cir] at (21,0) (aF) {$\{w(d_1)\}$};%

 \node[cir] at (23,0) (aG) {$\{r(d_2)\}$};%

 \node[cir] at (25,0) (aH) {$\{w(d_1)\}$};%

 \node[cir] at (27,0) (aI) {$\{r(d_2)\}$};%

 \node[cir] at (29,0) (aJ) {$\{r(d_1)\}$};%

 \node[cir] at (31,0) (aK) {$\{\nop\}$};

 \node[cir] at (33,0) (aL) {$\{r(d_1)\}$};%

 \path (aA) edge node {}node {} (aB);
 \path (aB) edge node {}node {} (aC);
 \path (aC) edge node {}node {} (aD);
 \path (aD) edge node {}node {} (aE);
 \path (aE) edge node {}node {} (aF);
 \path (aF) edge node {}node {} (aG);
 \path (aG) edge node {}node {} (aH);
 \path (aH) edge node {}node {} (aI);
 \path (aI) edge node {}node {} (aJ);
 \path (aJ) edge node {}node {} (aK);
 \path (aK) edge node {}node {} (aL);

 \path(aB) edge[bend left=25] node[above] {$d_1$} node{}(aE);
 \path(aF) edge[bend left=20] node[above] {$d_1$} node{}(aL);
 \path(aH) edge[bend left=25] node[above] {$d_1$} node{}(aJ); 

 \path(aC) edge[bend left=25] node[above] {$d_2$} node{}(aG);
 \path(aD) edge[bend left=20] node[above] {$d_2$} node{}(aI); 

 \end{tikzpicture}
}
 \caption{A valid sequence $\sigma=\nop ~w(d_1)~ w(d_2)~ w(d_2) ~r(d_1) ~w(d_1) ~r(d_2)~w(d_1)~r(d_2) ~r(d_1)~ \nop ~r(d_1)$ of operations from a system having two data structures
 (a stack $d_1$ and a queue $d_2$), with its graph $G_\sigma$.}
 \label{fig:untimed-data}
\end{figure}
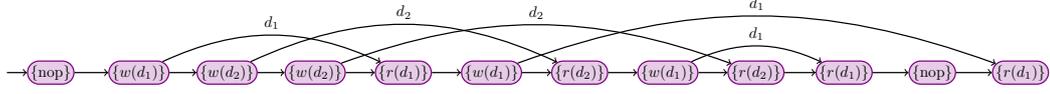
Let us fix a finite set of data structures $\DS$. Each data structure
$d\in \DS$ can be operated via two instructions, either a \emph{write} that
writes to the data structure, denoted $w(d)$, or a \emph{read}
instruction that reads from the data structure, denoted $r(d)$. The
set of instructions from $\DS$ is denoted $\Sigma^\DS=\{r(d),w(d)\myst d\in
\DS\}\cup \{\nop\}$, where $\nop$ is a special operation that does not access
the data-structures. For simplicity and ease of exposition, we restrict each
$d\in\DS$ to be a stack or a queue. However, the approach described here
can be adapted to other structures (such as bags) with minor modifications.
When $d\in \DS$ is a stack, $r(d)$ is the $\pop$ operation and $w(d)$ is the
$\push$ operation on stack $d$. Similarly, if $d$ is a queue, $r(d)$ is
the dequeue operation, while $w(d)$ is the enqueue operation on queue $d$.

A sequence of operations from $\Sigma^{\DS}$ abstracts a run of a system with
these data structures. We can then define the system as a generator of
(possibly infinitely many) sequences of operations.
The mechanism for generating this sequence of operations
can be some machine (an automaton), or can be specified
by regular expressions. We do not dwell on this detail here, and instead
 define \emph{a system $\cS$ with data
structures} as a regular language of sequences of operations over $\Sigma^\DS$.
Without loss of generality, we assume that all sequences will
start with $\nop$. It is easy to see
that standard models such as (multi)pushdown automata,
(multi)queue automata, multiset automata and so on generate regular languages of sequences of such
operations.

A sequence $\sigma$ of operations over $\Sigma^\DS$ is said to be
\emph{valid} if, for every prefix $\sigma'$ of $\sigma$ and for every data
structure $d\in\DS$, the number of reads $r(d)$ in $\sigma'$ is at most the
number of writes $w(d)$ in $\sigma'$, and the number of reads and writes in
$\sigma$ are equal. For a system $\cS$, we are only interested in
\emph{valid} sequences generated by $\cS$, and we denote this set by $L(\cS)$.
For instance, a valid behavior of a pushdown system cannot read/pop from a
stack before writing/pushing to it.
Let $\Gamma^\DS=\DS\cup \{\Succ\}$.
We associate, to any valid sequence $\sigma$ of operations over $\Sigma^\DS$, a $(\Sigma^\DS,\Gamma^\DS)$ linear graph $G_{\sigma}$.
\begin{definition}
 Let $\sigma=\sigma_1\ldots \sigma_n$ be a valid sequence of operations over
 $\Sigma^\DS$. We define its $(\Sigma^\DS,\Gamma^\DS)$-graph as
 $G_\sigma=(V,E,\lambda)$, where $V=\{1,\ldots n\}$ and
 \begin{enumerate}[nosep]
 \item for $1\leq i\leq n$, $\lambda(i)=\{\sigma_i\}$, and, for $1{\leq} i {< }n$, $i \xrightarrow{\Succ} i+1$,
 \item $\sigma_i=w(d)$ ($r(d)$) iff there is an outgoing (incoming) edge in $E$ labeled $d$ from (to) $i$.
 \item for each stack (queue) $d$, edges labeled $d$ satisfy the LIFO (FIFO) property.

 \end{enumerate}
\end{definition}
As an example, let $\sigma$ be a
sequence of operations from $\DS=\{d_1,d_2\}$, where $d_1$ is a stack and $d_2$
is a queue. The graph $G_\sigma$ corresponding to $\sigma$ is depicted in Figure~\ref{fig:untimed-data},
where the node labels are exactly the singleton sets of operations $w(d)$ and $r(d)$, for
$d \in \{d_1, d_2\}$.
We remark that this graph depends crucially on the interpretation of the data
structure, as a stack or a queue. Notice that the edges labeled $d_1$ respect the stack discipline (well-nesting), while the edges labeled $d_2$ respect FIFO. For a fixed $\DS$, we assume the interpretation of each data
structure to be fixed and simply write $G_\sigma$.

Given a $(\Sigma,\Gamma^\DS)$-graph $G=(V,E,\lambda)$, we define its
projection $\pi(G)$ as the $(\emptyset,\Gamma^{\DS})$-graph obtained by
removing the node labels: $\pi(G)=(V,E)$.

\begin{theorem}[\cite{mpri-lect}]
\label{thm:untimed}
 Let $\cS$ be a system with data structures from $\DS$. We can construct an
 $\EQLCPDL(\emptyset,\Gamma^{\DS})$ formula $\psi_\cS$ such that, for all
 $(\emptyset,\Gamma^{\DS})$-graphs $G$, $G\models \psi_\cS$ iff
 $G=\pi(G_\sigma)$ for some $\sigma\in L(\cS)$.
\end{theorem}

The classical \emph{non-emptiness problem} for a system $\cS$ with data
structures can be formulated as whether $L(\cS)\neq\emptyset$.

\begin{corollary}
 For system $\cS$, $\psi_\cS$ is satisfiable iff $L(\cS)\neq \emptyset$.
\end{corollary}

This corollary, along with Theorem~\ref{thm:pdl-sat}, and using known bounds on
tree-width, provides a ``uniform'' proof for the decidability of checking non-emptiness for a variety of untimed systems including (multi)pushdown
and (multi)queue systems
with bounded contexts, scope, or phases in a sequential setting. In many cases, the complexity obtained matches the best known bounds. We extend this
approach uniformly to timed systems, using the
realizability proof of Section~\ref{sec:realizability}.

\begin{figure}[t]
\begin{center}
  \scalebox{0.65}{
  \begin{tikzpicture}[->,thick] 
 \node[state, draw=white, rounded rectangle,minimum size=1.5em,inner sep=0em] at (-1,-1.4) (B0) {$\tau=$};

 \node[initial, cir, initial text ={}] at (0,0) (aA) {$1$} ;
 \draw[dashdotted,-] (0,-0.4)--(0,-1.1);
 \node[state, draw=white, rounded rectangle] at (0,-1.8) (B) {$\begin{array}{c} \nop \\ x:=0\\y:=0 \end{array}$};

 \node[cir] at (1.5,0) (aB) {$2$};
 \draw[dashdotted,-] (1.5,-0.4)--(1.5,-1.1);
 \node[state, draw=white, rounded rectangle] at (1.5,-1.6) (B1) {$\begin{array}{c} w(d) \\ \textcolor{red}{x=0} \\ \end{array}$};

 \node[cir] at (3,0) (aC) {$3$};
 \draw[dashdotted,-] (3,-0.4)--(3,-1.1);
 \node[state, draw=white, rounded rectangle] at (3,-1.6) (B2) {$\begin{array}{c} \nop \\ y:=0 \\ \end{array}$};

 \node[cir] at (4.5,0) (aD) {$4$};
 \draw[dashdotted,-] (4.5,-0.4)--(4.5,-1.1);
 \node[state, draw=white, rounded rectangle] at (4.5,-1.6) (B3) {$\begin{array}{c} w(d) \\ \textcolor{blue}{y \leq 1}\\ \end{array}$};

 \node[cir] at (6.3,0) (aE) {$5$};
 \draw[dashdotted,-] (6.3,-0.4)--(6.3,-1.1);
 \node[state, draw=white, rounded rectangle] at (6.3,-1.6) (B4) {$\begin{array}{c} r(d) \\ \textcolor{violet}{2<d-y} \end{array}$};

 \node[cir] at (7.9,0) (aF) {$6$};%
 \draw[dashdotted,-] (7.9,-0.4)--(7.9,-1.1);
 \node[state, draw=white, rounded rectangle] at (7.9,-1.6) (B5) {$\begin{array}{c} \nop \\ x:=0 \\ \end{array}$};

 \node[cir] at (9.5,0) (aG) {$7$};%
 \draw[dashdotted,-] (9.5,-0.4)--(9.5,-1.1);
 \node[state, draw=white, rounded rectangle] at (9.5,-1.4) (B6) {$\begin{array}{c} w(d) \end{array}$};

 \node[cir] at (11.5,0) (aH) {$8$};%
 \draw[dashdotted,-] (11.5,-0.4)--(11.5,-1.1);
 \node[state, draw=white, rounded rectangle] at (11.5,-1.8) (B7) {$\begin{array}{c} r(d) \\ \fole d \leqfi \\
 \textcolor{violet!50!yellow}{2 \leq x}\end{array}$};

 \node[cir] at (13.5,0) (aI) {$9$};%
 \draw[dashdotted,-] (13.5,-0.4)--(13.5,-1.1);
 \node[state, draw=white, rounded rectangle] at (13.5,-1.6) (B8) {$\begin{array}{c} \nop \\ \textcolor{purple!70!white}{y-x<6} \\ \end{array}$};

 \node[cir] at (15.5,0) (aJ) {$10$};%
 \draw[dashdotted,-] (15.5,-0.4)--(15.5,-1.1);
 \node[state, draw=white, rounded rectangle] at (15.5,-1.6) (B9) {$\begin{array}{c} r(d) \\ \textcolor{green!80!blue}{x-d<3} \\ \end{array}$};

 \path (aA) edge node {}node {} (aB);
 \path (aB) edge node {}node {} (aC);
 \path (aC) edge node {}node {} (aD);
 \path (aD) edge node {}node {} (aE);
 \path (aE) edge node {}node {} (aF);
 \path (aF) edge node {}node {} (aG);
 \path (aG) edge node {}node {} (aH);
 \path (aH) edge node {}node {} (aI);
 \path (aI) edge node {}node {} (aJ);

 \path(aB) edge[bend left=25] node[above] {$d$} node{}(aE);
 \path(aD) edge[bend left=25] node[above] {$d$} node{}(aH);
 \path(aG) edge[bend left=25] node[above] {$d$} node{}(aJ); 

  \end{tikzpicture}
}
\end{center}

\caption{A labeled linear graph $G_{\tau}$ obtained from a sequence of instructions $\tau$ from $\Sigma^{\DS}_{\clocks}$. For readability, the nodes are numbered and their instruction labels are written below them.}
 \label{fig:timed-data-a}
\end{figure}
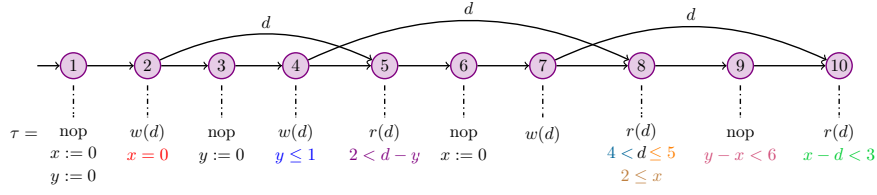

\begin{figure}[t]
  \begin{center}
  \scalebox{0.75}{
  \begin{tikzpicture}[->,thick] 
 \node[initial, cir, initial text ={}] at (17,-1) (xA) {$1$} ;

 \node[cir] at (18,-1) (xB) {$2$};
 \node[cir] at (19,-1) (xC) {$3$};
 \node[cir] at (20,-1) (xD) {$4$};
 \node[cir] at (21,-1) (xE) {$5$};
 \node[cir] at (22,-1) (xF) {$6$};%
 \node[cir] at (23,-1) (xG) {$7$};%
 \node[cir] at (24,-1) (xH) {$8$};%
 \node[cir] at (25,-1) (xI) {$9$};%
 \node[cir] at (26,-1) (xJ) {$10$};%

 \path (xA) edge node {}node {} (xB);
 \path (xB) edge node {}node {} (xC);
 \path (xC) edge node {}node {} (xD);
 \path (xD) edge node {}node {} (xE);
 \path (xE) edge node {}node {} (xF);
 \path (xF) edge node {}node {} (xG);
 \path (xG) edge node {}node {} (xH);
 \path (xH) edge node {}node {} (xI);
 \path (xI) edge node {}node {} (xJ);

 \path(xA) edge[draw=red,bend left=40] node[above] {\textcolor{red}{$\leq 0$}} node{}(xB);
 \path(xB) edge[draw=red,bend left=40] node[below] {\textcolor{red}{$\leq 0$}} node{}(xA);
 \path(xC) edge[draw=violet,bend left=40] node[below] {\textcolor{violet}{$< -2$}} node{}(xB);
 \path(xC) edge[draw=blue,bend left=40] node[above] {\textcolor{blue}{$\leq 1$}} node{}(xD);
 \path(xC) edge[draw=purple!70!white,bend left=80] node[above] {\textcolor{purple!70!white}{$< 6$}} node{}(xF);

 \path(xD) edge[draw=red!50!yellow,bend left=70] node[above] {\textcolor{red!50!yellow}{$\leq 5$}} node{}(xH);
 \path(xH) edge[draw=blue!60!green,bend left=60] node[below] {\textcolor{blue!60!green}{$< -4$}} node{}(xD);

 \path(xF) edge[draw=green!80!blue, bend left=40] node[above] {\textcolor{green!80!blue}{$< 3$}} node{}(xG);
 \path(xH) edge[draw=violet!50!yellow,bend left=30] node[below] {\textcolor{violet!50!yellow}{$\leq -2$}} node{}(xF);

 \end{tikzpicture}
  }
   \end{center}
  \caption{
    The weighted graph $\wg_\tau$ corresponding to the sequence of instructions $\tau$ (from Figure~\ref{fig:timed-data-a}).}
 \label{fig:timed-data-b}
\end{figure}
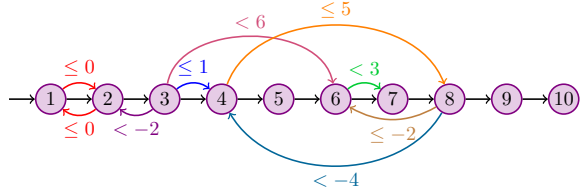

\subsection{Combining timing and data structures}\label{sec:DS+time}

While combining time constraints and data structures, we cannot directly rely
on the formula for realizability from Section~\ref{sec:realizability} in the approach outlined above. The vocabulary of graphs obtained from systems having time constraints and data structures might differ from the (weighted) $(\emptyset,\Gamma^{M})$-graphs of Section~\ref{sec:realizability} and the (unweighted) $(\Sigma,\Gamma^{\DS})$-graphs above, where $\Sigma=\emptyset$ or
$\Sigma=\Sigma^{\DS}$.
The crucial observation is that, for a large class of timing constraints and
data structures that we are interested in, it turns out that the former weighted graphs
can be interpreted in the latter unweighted graphs, paving the way to extend the approach for systems having both time constraints and data structures. We now detail this intuition.

\subsubsection{Timing instructions}
\label{sec:ti}
In a timed system with data structures, the sequence of instructions generated
by the system includes (i) checking time constraints on clocks (encoded as operations on clocks),
(ii) checking time constraints on data structures, and (iii) mixing operations on clocks and data structures.
Recall that we already have a fixed set of data structures $\DS$ consisting of
stacks and queues. To be concrete, we also fix a representative set of timing
features.

We fix a finite set $\clocks$ of real-valued ``clock'' variables and a maximal
constant $M\in \mathbb{N}$. We also fix notations
$\bow\in{\{\leq,<,=,>,\geq\}}$, $\beta\in[0,M)\cap\mathbb{N}$ and use
letters $x,y,x_1,\ldots $ for clock variables. Atomic timing instructions are as follows:
\begin{enumerate}
 \item for $x\in \clocks$, $x{:=}0$ represents \emph{clock resets}, while $x\bow \beta$
 represent \emph{guards} or \emph{clock constraints};

 \item for $d\in \DS$, $d\bow \beta$ represents
 an \emph{age constraint} checking the ``age'' of the message read;

 \item for $d\in \DS$ and $x,y\in \clocks$,
 $(x-y)\bow \beta$, $(d-x)\bow \beta$ and $(x-d)\bow \beta$ represent \emph{diagonal constraints}.
 The latter two capture mixing clock variables and data structures.
 \end{enumerate}

Thus, we define a set of instructions $\Sigma^{\DS}_{\clocks}$ which contains
$\Sigma^{\DS}$ with the atomic timing instructions described above.
Without loss of generality, we only consider sequences of instruction sets (also called \emph{sequences of
instructions} for simplicity) from
$\Sigma^{\DS}_{\clocks}$ starting with the set $\{\nop\}\cup\{x{:=0}\myst
x\in\clocks\}$, i.e., which resets all clocks at start-up.
A sequence $\tau$ of such instructions
is shown in Figure~\ref{fig:timed-data-a}. We associate to
every such sequence
$\tau$ a sequence of untimed instructions $\sigma_\tau$, obtained by ignoring the atomic timing instructions. Now we say $\tau$ is
valid if $\sigma_\tau$ is valid. Then, for every valid $\tau$, we can
immediately associate a $(\Sigma^{\DS}_{\clocks},\Gamma^\DS)$-labeled linear
graph $G_\tau$ by considering $G_{\sigma_\tau}$ and enriching its
node labels with the timing instructions. 

We define a timed system with data structures $\TS$ as a regular
language of sequences of instructions over $\Sigma^{\DS}_{\clocks}$. It
is easy to see that classical models, such as timed automata, (multi-stack) timed pushdown automata or timed automata with gap order constraints, can be modeled in this formalism. The set of valid sequences generated by $\TS$ is denoted $L(\TS)$. Now, a valid sequence of instructions $\tau=\tau_1\ldots\tau_n$ over
$\Sigma^{\DS}_{\clocks}$ is said to be \emph{timed feasible} or just
\emph{feasible} if there exists a time-stamping
$\ts:\{1,\ldots,n\}\rightarrow \mathbb{R}^{\geq 0}$ such that all timing
constraints engendered by the timing instructions are satisfied. That is, for
$\bow\in \{\leq,<,=,>,\geq\}$ and $\beta\in \mathbb{N}$:
\begin{enumerate}[nosep,label=($\mathsf{C}_{\arabic*}$),ref=$\mathsf{C}_{\arabic*}$]
 \item\label{item:C1} For every guard of the form $x \bow \beta$ at position
 $i$, if the last reset instruction of the clock $x$ in $\tau$ before $i$ was
 at position $j$, then $\ts(i)-\ts(j) \bow \beta$.

 \item\label{item:C2} For every age constraint of the form $d \bow \beta$ at position
 $i$, we have an edge $j\Edge{d}i$ in $G_\tau$ (which implies
 $w(d)\in\lambda(j)$ and $r(d)\in\lambda(i)$),
 and $\ts(i)-\ts(j) \bow \beta$.

 \item\label{item:C3} For every diagonal constraint of the form $x-y \bow\beta$ at
 position $i$, if $j$ and $k$ are the last resets of clocks $x$ and $y$
 respectively, then $\ts(k)-\ts(j) \bow\beta$.

 \item\label{item:C4} We can similarly define
 diagonal constraints between clocks and data structures.
\end{enumerate}
Thus, the \emph{non-emptiness problem} for the timed system $\TS$ is to check
whether there exists a feasible $\tau\in L(\TS)$. 

\subsubsection{From timing instructions to weighted graphs}
We reduce checking non-emptiness of $\TS$ to checking satisfiability of an \pdl formula over $(\Sigma^{\DS}_{\clocks},\Gamma^{\DS})$-graphs. Towards this, we first define the weighted graph $\wg_\tau$ corresponding to a valid sequence of instructions $\tau$ of $\TS$ in a
natural manner. We extend from Section~\ref{sec:realizability}, where all timing instructions were simply clock constraints
and resets of clocks i.e., corresponding to \eqref{item:C1} and \eqref{item:C3} above. In Figure~\ref{fig:timed-data-a}, the check of $\textcolor{red}{x=0}$ on node 2 gives two bidirectional weighted edges in the weighted graph $\wg_\tau$ depicted in Figure~\ref{fig:timed-data-b}, between the last reset point of $x$ and node 2. Similarly, instruction $\textcolor{blue}{y\leq 1}$ at node 4 gives rise to the forward edge labeled $\textcolor{blue}{\leq 1}$ between last reset of $y$ and node 4. For diagonal constraints \eqref{item:C3}, the edge obtained is between the last reset points. E.g, $\textcolor{purple!70!white}{y-x<6}$ at node 9 yields the weighted edge from node 3 to node 6 (last resets of clocks $y$ and $x$).

This construction easily lifts to \eqref{item:C2} and \eqref{item:C4} as well. For \eqref{item:C2}, we just observe that each age constraint engenders edges between the source write and target read of that data structure edge. E.g., in Figure~\ref{fig:timed-data-a}, the age constraint $\fole d \leqfi$
at node 8 yields two weighted edges (in Figure~\ref{fig:timed-data-b}) between the source of the data structure edge, i.e., node 4 and target, node 8. The upper bound is captured by the
forward edge while the lower bound by the backward edge.
Similarly the constraint $\textcolor{violet}{2<d-y}$ at node 5 yields the backward edge from node 3 (the last reset of clock $y$) to node 2 (the source of the data structure edge reaching node 5) labeled $\textcolor{violet}{<-2}$ (as it is a lower bound constraint).

The main property about the weighted graph is that it captures feasibility of a sequence of instructions as realizability.
\begin{lemma}
 A valid sequence of instructions $\tau$ over $\Sigma^{\DS}_{\clocks}$ is
 feasible iff $\wg_\tau$ is realizable.
\end{lemma}

\subsubsection{Interpreting weighted graphs in unweighted graphs}
From the above discussion, given a timed system $\TS$, for each valid $\tau$ of $\TS$, we have a weighted
graph $\wg_\tau$. A significant contribution of this paper, of possible independent interest, is the following proposition which relates these weighted graphs with unweighted $(\Sigma^{\DS}_{\clocks},\Gamma^{\DS})$-graphs obtained from $\tau$. 
Proposition \ref{prop:graphinterpret} allows us to logically interpret weighted graphs into unweighted ones and, therefore, to decouple the data structure and process edges from the timing constraints.

\begin{proposition}
 \label{prop:graphinterpret}
 Let $\tau$ be a valid sequence of instructions over $\Sigma^{\DS}_{\clocks}$.
 Then the weighted graph $\wg_\tau$ can be $\CPDL$-interpreted in the
 $(\Sigma^{\DS}_{\clocks},\Gamma^{\DS})$-graph $G_\tau$.
\end{proposition}
\begin{proof}
 Given a valid sequence of instructions $\tau$ over $\Sigma^{\DS}_{\clocks}$,
 let $M$ be the maximal constant appearing in these instructions. We saw in
 the previous subsection that the weighted graph $\wg_\tau=(V,E)$ has successor
 edges, and weighted edges arising from constraints of type
 (\ref{item:C1}--\ref{item:C4}).
 First, we observe that successor edges in $\wg_\tau$ are already present as
 successor edges in $G_\tau$. For weighted edges, let ${\lleq}\in\{<,\leq\}$,
 and $c\in [0,M)\cap \mathbb{N}$.
 We assume that equality constraints such as $x=c$ have been replaced by the
 conjunction of $x\leq c$ and $c\leq x$.
 For a clock $x\in\clocks$, we define the path formula\\
 $\creset_x = {\proc}^{-1}\cdot(\test{\neg\reset{x}}\cdot{\proc}^{-1})^{*}\cdot\test{\reset{x}}$\\
 which moves backwards along successor edges up to the last reset of clock $x$.
 Then, towards the interpretation of forward edges weighted with ${}\lleq c$,
 we define the path formula $\Pi_{\lleq c}$ as
 \begin{align*}
 &\hspace{1mm} \sum_{x\in\clocks} \hspace{1mm}
 \creset_x^{-1}\cdot\test{x\lleq c}
 \tag{\ref{item:C1}}
 \\
 +&\hspace{3mm} \sum_{d\in\DS} \hspace{3mm}
 \Edge{d}\cdot\test{d\lleq c}
 \tag{\ref{item:C2}}
 \\
 +& \sum_{x,y\in\clocks} \creset_x^{-1}\cdot\test{x-y\lleq c}\cdot\creset_y
 \tag{\ref{item:C3}}
 \\
 +&\hspace{1mm} \sum_{\substack{x\in\clocks\\ d\in\DS}} \hspace{1mm}
 \creset_x^{-1}\cdot\test{x-d\lleq c}\cdot\Edgem{d}
 \tag{\ref{item:C4}}
 \\[-6mm]
 &\hspace{18mm}
 {}+\Edge{d}\cdot\test{d-x\lleq c}\cdot\creset_x
 \end{align*}
 Then, for all $u,v\in V$ and $c>0$ (we will discuss the case $c=0$ below), we
 have $(u,\lleq,c,v)\in E$ iff $(G_\tau,u,v)\models \Pi_{\lleq c}$. The four
 types of \emph{upper} constraints defined in (\ref{item:C1}--\ref{item:C4})
 are described by the respective path formulae (\ref{item:C1}--\ref{item:C4})
 in $\Pi_{\lleq c}$. As an example, if we refer to the $i$\textsuperscript{th} node of $G_{\tau}$ as $u_i$ in Figure \ref{fig:timed-data-a} and $\wg_\tau$ in Figure~\ref{fig:timed-data-b}, we have
 the edge $(u_3, \lleq, 6, u_6)$ in $\wg_\tau$ because $(G_\tau,u_3,u_6)\models \creset_y^{-1}\cdot\test{y-x\lleq 6}\cdot\creset_x$.
 Similarly, the edge $(u_6, <, 3, u_7)$ is present in $\wg_\tau$
 since $(G_{\tau}, u_6,u_7) \models \creset_x^{-1}\cdot\test{x-d < 3}\cdot\Edgem{d}
 $.
 Notice that in $\creset_x$, we walk backward to the \emph{first} node labeled $x:=0$, while, in
 \ref{item:C2} and \ref{item:C4}, for checking the age of a data structure, it is sufficient
 to check the existence of a data structure backward edge from the point where the age is checked.

 Similarly, towards the interpretation of backward edges weighted with
 ${}\lleq -c$, we define the path formula $\Pi_{\lleq -c}$
 as
 \begin{align*}
 &\hspace{1mm} \sum_{x\in\clocks} \hspace{1mm}
 \test{c\lleq x}\cdot\creset_x
 \tag{\ref{item:C1}}
 \\
 +&\hspace{3mm} \sum_{d\in\DS} \hspace{3mm}
 \test{c\lleq d}\cdot\Edgem{d}
 \tag{\ref{item:C2}}
 \\
 +& \sum_{x,y\in\clocks} \creset_y^{-1}\cdot\test{c\lleq x-y}\cdot\creset_x
 \tag{\ref{item:C3}}
 \\
 +&\hspace{1mm} \sum_{\substack{x\in\clocks\\ d\in\DS}} \hspace{1mm}
 \creset_x^{-1}\cdot\test{c\lleq d-x}\cdot\Edgem{d}
 \tag{\ref{item:C4}}
 \\[-6mm]
 &\hspace{18mm}
 {}+\Edge{d}\cdot\test{c\lleq x-d}\cdot\creset_x
 \end{align*}
 Then, for all $u,v\in V$ and $c>0$, we have $(u,\lleq,-c,v)\in E$ iff
 $(G_\tau,u,v)\models \Pi_{\lleq -c}$. Again, the four types of \emph{lower}
 constraints defined in (\ref{item:C1}--\ref{item:C4}) are described by the
 respective path formulae (\ref{item:C1}--\ref{item:C4}) in $\Pi_{\lleq -c}$.

 Now, when $c=0$, an edge weighted $\lleq 0$ may arise from an upper constraint
 such has $x\lleq 0$ or a lower constraint such as $0\lleq x$. Therefore, for all
 $u,v\in V$, we have $(u,\lleq,0,v)\in E$ iff $(G_\tau,u,v)\models \Pi_{\lleq 0}
 + \Pi_{\lleq -0}$.

The size of $\Pi_{\lleq\alpha}$ is
 $\mathcal{O}(|\clocks|^{2}+|\DS|+|\clocks||DS|)$.

 Thus we have described how each edge of the weighted graph $\wg_\tau$ can be
 interpreted in the $(\Sigma^{\DS}_{\clocks},\Gamma^{\DS})$-graph $G_\tau$ by
 an $\CPDL$-formula,
 of size $\mathcal{O}(|\clocks|^{2}+|\DS|+|\clocks||DS|)$,
 which
 completes the proof of this proposition.
\end{proof}
Thus, any formula over weighted graphs can be translated
into an ``\emph{equivalent}'' formula over
$(\Sigma^{\DS}_{\clocks},\Gamma^{\DS})$-graphs:

\newcommand{\SigmaDSC}{\Sigma^\DS_\clocks}
\newcommand{\GammaDS}{\Gamma^\DS}

\begin{corollary}
 \label{cor:backtrans}
 Given a formula $\psi\in \pdl(\emptyset, \Gamma_M)$, we can construct
 $\psi'\in\pdl(\Sigma^\DS_{\clocks},\Gamma^\DS)$ such that, for all valid
 sequences of instructions $\tau$ over $\Sigma^{\DS}_{\clocks}$, we have
 $\wg_\tau \models \psi$ iff $G_\tau \models \psi'$. The size of $\psi'$ is
 $\mathcal{O}((|\clocks|^{2}+|\DS|+|\clocks||DS|)|\psi|)$ and its intersection
 width is same as $\psi$.
\end{corollary}

\subsubsection{Reducing emptiness of $\TS$ to satisfiability of $\pdl$}

From Theorem~\ref{thm:linear}, we know that there exists a formula capturing
realizability on weighted graphs, with signature $(\emptyset,\Gamma_M)$.
Combining with Corollary~\ref{cor:backtrans} gives us the second main theorem of
the paper regarding logical characterization of emptiness checking in timed systems with data structures.
\begin{theorem}[Logical characterization of a timed system]
 \label{thm:timed}
 Given a timed system with data structures $\TS$, we can construct a formula
 $\Psi_\TS\in\pdl(\emptyset,\Gamma^\DS)$ such that for all
 $(\emptyset,\Gamma^{\DS})$ linear graphs $G$, we have $G\models \Psi_\TS$ iff
 $G=\pi(G_\tau)$ for some \emph{feasible} $\tau\in L(\TS)$.
 The size of $\Psi_\TS$ is polynomial in the size of $\TS$
 and its intersection width is 2.

\end{theorem}
\begin{proof}
 By Theorem~\ref{thm:linear}, we can construct a formula $\mathsf{Realizable}$
 in $\pdl(\emptyset,\Gamma_M)$ that captures realizability over weighted graphs
 $\TCG$. By Corollary~\ref{cor:backtrans}, we obtain a formula
 $\psi_{\mathsf{real}}\in\pdl(\emptyset,\Gamma^\DS)$ such that, for all $\tau\in
 L(\TS)$, $G_\tau\models\psi_{\mathsf{real}}$ iff
 $\wg_\tau\models\mathsf{Realizable}$. In fact, $\psi_{\mathsf{real}}$ is
 simply obtained from $\mathsf{Realizable}$ by replacing every reference to a
 weighted edge in the formula by its logical interpretation in $G_\tau$. Now,
 by definition of $\pdl$, we have $\psi_{\mathsf{real}}=\exists p_1\ldots p_r
 \psi'$ for some $\psi'\in \ICPDL(\{p_1,\ldots p_r\},\Gamma^\DS)$.

 Next, recall that a timed system $\TS$ is a regular language of sequences of
 timed instructions. We consider the automaton that describes this regular
 collection, denoted by $\cA=(Q,i,F,\Delta)$ with $Q$ the set of states, $i$
 the initial state and $F$ the final states and $\Delta$ the transition
 function. Then, the accepted sequences of instructions can be captured in
 $\EQLCPDL$, by guessing the states visited along an accepting run, and by checking
 that consecutive states have a transition between them and start from initial
 and end at final state. 
 Though similar in spirit to Theorem~\ref{thm:untimed}, for the sake of
 completeness and to obtain the precise complexity, we detail the construction
 in Appendix~\ref{app:proofoftheorem}.

 Set $\Sigma=\Sigma^\DS_\clocks\cup Q=\{q_1,\ldots,q_n\}$.
 There exists a formula $\xi=\exists q_1\ldots
 q_n \xi'$, with $\xi'\in \LCPDL(\Sigma,\Gamma^\DS)$, such that, for all
 $(\emptyset,\Gamma^\DS)$-graphs $G$, we have $G\models \xi$ iff
 $G=\pi(G_\tau)$ for some sequence $\tau\in L(\TS)$.
Combining this with the formula above, and define $\psi_\TS=\exists
 p_1\ldots p_r, q_1,\ldots q_n (\xi'\wedge\psi')$. Then we have for any
 $(\emptyset,\Gamma^\DS)$-graph $G$, $G\models \psi_\TS$ iff $G=\pi(G_\tau)$
 for some $\tau\in L(\TS)$ and $\tau$ is feasible, which completes the proof.
\end{proof}

\subsection{Application: deciding emptiness}
\label{sec:sat-complexity}

While we have reduced checking emptiness of timed systems to checking satisfiability of a formula in $\pdl$, this does not immediately give decidability results. This is obvious since systems with multiple data structures (such as stacks or even single queue) are all Turing powerful, even without any timing features. To obtain decidability, one often considers under-approximations, for which we essentially restrict the class of graphs that are considered as behaviors. As mentioned in the preliminaries, graphs of bounded tree-width form a large family of graphs where we regain decidability thanks to Theorem~\ref{thm:pdl-sat}. Recall that $\Graphs^k$ denotes graphs of tree-width at most $k$. Combining Theorems~\ref{thm:pdl-sat} and~\ref{thm:timed}, we have the following corollary about decidability in timed systems. 

\begin{corollary}[Underapproximations.]\label{cor:decidability}
 Let $k\in \mathbb{N}$. Let $\cS$ be a timed system with data structures that uses clocks from $\clocks$ and has maximum constant $M\in\mathbb{N}$. We can check whether there exists a feasible $\tau\in L(\cS)$ such that $G_\tau \in \Graphs^k(\emptyset,\Gamma^\DS)$ in time $2^{\poly(k,M,|\clocks|,|\DS|)}\times|\cS|^{\poly(k,|\DS|)}$.
 \end{corollary}
Thus, if the set $\{G_\tau\myst \tau\in L(\cS)\}$ has a bounded tree-width, we obtain the same complexity bounds for checking emptiness of $\cS$. As concrete applications, the following models of timed systems all fall in this category of having bounded tree-width, hence we obtain decidability (and efficient algorithms) for checking emptiness of
timed automata~\cite{AD94}, dense-timed pushdown automata with a single stack~\cite{lics12}, multi-stack dense-timed pushdown automata with bounded rounds~\cite{concur16}.
In fact, the complexity obtained for dense-timed pushdown automata with a single stack is even optimal.
In addition, by this technique, we also have the following (new, to the best of our knowledge) results on the decidability of the emptiness
problem for multi-stack dense-timed pushdown automata with
 (i) bounded contexts (the tree-width of graphs in the case of $p$-bounded context systems is $\leq p+1$~\cite{MP11}),
 (ii) bounded phase (the tree-width of graphs in the case of $p$-bounded phase systems is $\leq 2^{p+1}$~\cite{CGK12}), and
 (iii) bounded scope (the tree-width of graphs in the case of $p$-bounded scope is $\leq 2(p+2)$~\cite{CGK12}).
Further, if one considers timed automata with $b$-bounded channels (a $b$-bounded channel is one where the number of unread messages
is bounded by $b \in \mathbb{N}$ at any point of time), then the $(\emptyset,\Gamma^{\DS})$-graphs have a tree-width $\leq b+2$~\cite{mpri-lect}. We expect that many other data structures and various novel combinations (e.g., any combination of the above with multiple stacks and queues) can be handled using our technique, and leave these as routine exercises. 
In the next section, we consider more substantial extensions.
\begin{figure*}[t]
 \begin{center}
 \scalebox{0.6}{
 \begin{tikzpicture}[->,thick]
 \node[state, draw=white, rounded rectangle,minimum size=1.5em,inner sep=0em] at (-1.5,-1.4) (B0) {$\tau=$};
 \node[initial, cir, initial text ={}] at (0,0) (aA) {} ;
 \draw[dashdotted,-] (0,-0.4)--(0,-1.1);
 \node[state, draw=white, rounded rectangle] at (0,-1.8) (B) {$\begin{array}{c} x_1:=0\\x_2:=0\\x_3:=0 \end{array}$};
 \node[cir] at (2,0) (aB) {};
 \draw[dashdotted,-] (2,-0.4)--(2,-1.1);
 \node[state, draw=white, rounded rectangle] at (2,-1.6) (B1) {$\begin{array}{c} d_1:=x_1\\ x_2:=0 \end{array}$};
 \node[cir] at (4,0) (aC) {};
 \draw[dashdotted,-] (4,-0.4)--(4,-1.1);
 \node[state, draw=white, rounded rectangle] at (4,-1.6) (B2) {$\begin{array}{c} x_2:=d_1 \\x_1:=0\\x_4:=x_2 \end{array}$};
 \node[cir] at (6,0) (aD) {};
 \draw[dashdotted,-] (6,-0.4)--(6,-1.1);
 \node[state, draw=white, rounded rectangle] at (6,-1.6) (B3) {$\begin{array}{c} d_2:=x_2\\ x_2:=0 \\ x_3:=x_4\end{array}$};
 \node[cir] at (8,0) (aE) {};
 \draw[dashdotted,-] (8,-0.4)--(8,-1.1);
 \node[state, draw=white, rounded rectangle] at (8,-1.6) (B4) {$\begin{array}{c} x_4:=d_2\\ \textcolor{blue}{x_3<3} \end{array}$};
 \node[cir] at (10,0) (aF) {};%
 \draw[dashdotted,-] (10,-0.4)--(10,-1.1);
 \node[state, draw=white, rounded rectangle] at (10,-1.6) (B5) {$\begin{array}{c} \textcolor{red}{x_4<4} \end{array}$};
 \path (aA) edge node {}node {} (aB);
 \path (aB) edge node {}node {} (aC);
 \path (aC) edge node {}node {} (aD);
 \path (aD) edge node {}node {} (aE);
 \path (aE) edge node {}node {} (aF);
 \path(aB) edge[bend left=25] node[above] {$d_1$} node{}(aC);
 \path(aD) edge[bend left=25] node[above] {$d_2$} node{}(aE);

 \node[initial, cir, initial text ={}] at (12,-1) (xA) {} ;
 \node[cir] at (13.5,-1) (xB) {};
 \node[cir] at (15,-1) (xC) {};
 \node[cir] at (16.5,-1) (xD) {};
 \node[cir] at (18,-1) (xE) {};
 \node[cir] at (19.5,-1) (xF) {};

 \path (xA) edge node {}node {} (xB);
 \path (xB) edge node {}node {} (xC);
 \path (xC) edge node {}node {} (xD);
 \path (xD) edge node {}node {} (xE);
 \path (xE) edge node {}node {} (xF);

 \path(xB) edge[draw=blue,bend left=20] node[above] {\textcolor{blue}{$< 3$}} node{}(xE);
 \path(xA) edge[draw=red,bend left=25] node[above] {\textcolor{red}{$<4$}} node{}(xF);

 \end{tikzpicture}
 }
 \caption{Intricate flow of information in complex renamings.}
 \label{fig:complex}
 \end{center}
\end{figure*}
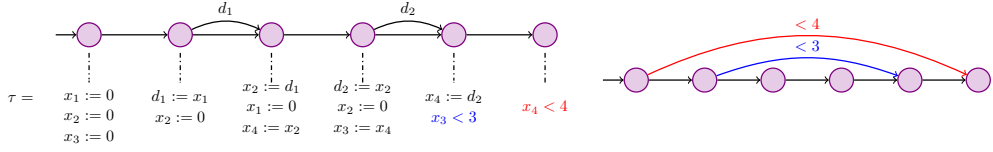

\section{Extensions}
\label{sec:ext}
We consider two extensions: first, adding new timing features without much change to the theory above, and second, extending from checking emptiness to model checking.
\subsection{Capturing time features - a generic template}
We develop a two-step template to add new timing features to our approach above.
Step 1 consists in expressing the edges engendered by the new feature in the
weighted graph and Step 2 consists in writing a formula in $\LCPDL$ to capture
this new edge relation. If we can accomplish these steps, then our theorems
lift to the setting with these new timing features.

This shows the robustness of our approach as we are able to handle these extra features, uniformly and with ease. At the same time, we remark that this template is interesting even for timing features that we know can be simulated by ordinary clocks. For instance, consider diagonal guards in timed automata, which are expressively equivalent to timed automata without diagonal guards. Removing diagonal constraints incurs an additional exponential blow-up in the worst-case~\cite{BC05}, which can be avoided by directly expressing their edges in the weighted graph as we did in Equation~\ref{item:C4}.

\subsubsection{Event clocks}
Let us illustrate this template in action via another example of a well-studied model, namely, event predicting clocks~\cite{AFH99,GRS14}, which can be simulated by ordinary (non-deterministic) timed automata. We fix a set $\AP$ of atomic propositions
(events) arising from the system.
An event-predicting timing
instruction $\nxt_a \bow \alpha$, for $a\in\AP$, $\bow\in
\{\leq,<,>,\geq\}$ and $\alpha\in [0,M)\cap\mathbb{N}$, entails a constraint between
the current point (call it $u$) and the point at which node label $a$ occurs
next (call it $v$). Consistently with the notations on timing constraints
\ref{item:C1}-\ref{item:C4},
in section \ref{sec:ti},
we call this constraint $\mathsf{C}_5$. Now, Step 1 is that this can be
expressed in the weighted graph as an edge between these two vertices $u$ and
$v$. For Step 2, it is easy to write the $\PDL$ formula that allows to
interpret these edges of the weighted graph as edges in the $\Gamma^\DS$-graph.
Specifically, we just have to add to the path formula $\Pi_{\lleq\alpha}$ in proof
of Proposition~\ref{prop:graphinterpret} the following term:
\begin{align*}
 \sum_{a\in \AP} \test{(\nxt_a\lleq \alpha)}\cdot \proc\cdot
 (\test{\neg a}\cdot \proc)^\ast\cdot \test{a}
 \tag{$\mathsf{C}_5$}\label{eq:eca-pdl}
\end{align*}
We proceed similarly for the path formula $\Pi_{\lleq-\alpha}$. It is not
difficult to see that we can define similar formulae to capture event recording
clocks as well.

\subsubsection{Clock renaming via tracking}
\label{sec:tracking}
While event clocks are relatively straightforward, for some other timing features, it is not easy to figure out, from the timing instruction, what edges in the weighted graph must be added. This happens for instance in clock renaming: if we assign to $x$ the value of clock $y$ and then check it later with $x\leq\alpha$, the edge to be added is from the last reset of $y$ to the point of checking the constraint. This is the case even if $y$ has been reset in between after the assignment. Figure \ref{fig:complex} illustrates this.

We consider a generic class of (deterministic) clock renaming in timed systems. Such renamings are a special case of clock updates, which are again a classical notion in timed automata~\cite{BDFP04,BC05}, but have not been studied much for timed systems with single or multiple data structures such as stacks and queues. We divide the features we consider into 4 classes:
\begin{itemize}
 \item[(i)] the usual reset of a clock $x$ to 0 ($x:=0$),
 \item[(ii)] assigning to clock $x$ the value of clock $x'$ ($x:=x'$),
 \item[(iii)] assigning to clock $x$ the value associated to data structure $d \in \DS$, while reading from $d$ ($x:=d$),
 \item[(iv)] writing to $d \in \DS$ the value of clock $x$ ($d:=x$).
\end{itemize}
Note that renamings (iii) and (iv), combined with the age and diagonal constraints on data structures, give us a rich class of timed systems. This allows us to consider timed systems where we can write to
some $d_1 \in \DS$ the value of a clock $x_1$, then read from $d_1$ this value
(which changes with passage of time) into a clock $x_2$, write this value of
$x_2$ to some $d_2 \in \DS$, and retrieve the value (after some time elapse)
into a clock $x_4$. This value in $x_4$ can then be checked with the value read
from some $d_4 \in \DS$, or with a clock $x_5$, or with a constant $\alpha$.
In such a sequence, the clock $x_1$ has come a long way at this time of checking, and we need to track it, to ensure that the time elapse we are looking at happens from the last reset of $x_1$ before it was written to $d_1$. See Figure \ref{fig:complex}, where the value of clock $x_1$ flows through $d_1, x_2, d_2$ and finally $x_4$, from where it is checked. Likewise, the value of clock $x_2$ flows through clocks $x_4, x_3$, and is checked
at $x_3$. Now, $x_2$ is reset after it flows into $x_4$; however, when checking $x_3$, we use the reset
of $x_2$ before $x_2$ flowed inside $x_4$. 

Many recent papers \cite{CL18,Clemente18,formats18} consider complex constraints between data structures and clocks; however, this intricate flow of information across clocks and data structures has not been looked at, to the best of our knowledge. In these papers, there are time constraints between data structures $d_1, d_2$, between clocks, and also between a clock $c$ and a data structure $d$. All of these can be modeled easily in our case, as shown in Figure \ref{fig:timed-data-a}. 

Inferring  constraints (i)-(iv) as above requires us to follow and track the clock reset back to the original event. Rather than writing a formula in $\CPDL$, we find it easier to describe an automaton which ``walks'' in the graph and performs this tracking. This enables us to express the weighted edges engendered by the
constraints using the accepting paths of the automaton. This essentially
handles the Step 1 we mentioned earlier.
To handle Step 2, which is the logical definability,
we write $\CPDL$ formulae whose paths $\pi$
use this automaton. This allows us to interpret the
weighted edges. 

Formally, we construct an automaton $\cA$ with set of states $Q=\{q_x\myst x\in \clocks\}$.
A run of $\cA$ starting from some state $q_x$ will track the name of the clock
whose value originates from $x$. Without loss of generality, we assume that
each transition of the timed system $\TS$ contains exactly one renaming operation for each
clock, which could be of the form $x:=0$ (reset), $x:=x'$ (deterministic clock
renaming, we use $x:=x$ if the clock is unchanged), $x:=d$ ($x$ is updated with
the value read from $d \in \DS$), or $d:=x$. There are two types of transitions:
\begin{itemize}[nosep]
 \item (clock renaming): if there is a renaming operation $x':=x$ then we have a
 transition $q_x\xrightarrow{\test{x':=x}\cdot{\to}}q_{x'}$,

 \item ($\DS$ renaming): if there is an renaming operation $x':=d$ for some $d \in
 \DS$, then for all clocks $x$, we have a transition
 $q_x\xrightarrow{\test{d:=x}\cdot{\Edge{d}}\cdot \test{x':=d}}q_{x'}$. This corresponds to writing the value of clock $x$ to some $d \in \DS$, and,
 at the time of reading from $d \in \DS$, assign this value to a clock $x'$.
\end{itemize}

Consider a run
$\rho=q_{x_0}\xrightarrow{\pi_1}q_{x_1}\xrightarrow{\pi_2}q_{x_2}\cdots\xrightarrow{\pi_n}q_{x_n}$
in $\cA$. Let $\tau\in L(\TS)$ be a valid sequence of instructions from the
timed system $\TS$. Let $G_\tau$ be the associated $(\SigmaDSC,\GammaDS)$-graph
and let $u,v$ be vertices in $G_\tau$. Then,
$G_\tau,u,v\models\mathsf{label}(\rho)=\pi_1\cdot\pi_2\cdots\pi_n$ iff the value
of clock $x_n$ at $v$ originates from clock $x_0$ at $u$. We write
$G_\tau,u,v\models\cA_{x,x'}$ if there is a run $\rho$ of $\cA$ from $q_x$ to
$q_{x'}$ such that $G_\tau,u,v\models\mathsf{label}(\rho)$.

Now, we can revisit and generalize the timing constraints above in
(\ref{item:C1}--\ref{item:C4}) using $\cA$ instead of the paths tracking the
last reset of a clock. For instance, the subformulae
(\ref{item:C1}--\ref{item:C3}) of $\Pi_{\lleq\alpha}$ in the proof of
Proposition~\ref{prop:graphinterpret} should be replaced with
\begin{align*}
 &\hspace{0mm} \sum_{x,x'\in\clocks} \hspace{-3mm}
 \test{\reset{x}}\cdot\cA_{x,x'}\cdot\test{x'\lleq\alpha}
 \tag{\ref{item:C1}}
 \\
 +&\hspace{0mm} \sum_{\substack{x,x'\in\clocks\\ d\in\DS}} \hspace{-3mm}
 \test{\reset{x}}\cdot\cA_{x,x'}\cdot\test{d:=x'}
 \\[-6mm]
 &\hspace{31mm}
 {}\cdot{\Edge{d}}\cdot\test{d\lleq\alpha}
 \tag{\ref{item:C2}}
 \\
 +& \hspace{-3mm}\sum_{x,x',y,y'\in\clocks} \hspace{-5mm}
 \test{\reset{x}}\cdot\cA_{x,x'}\cdot\test{x'-y'\lleq\alpha}
 \\[-4mm]
 &\hspace{31.5mm}
 {}\cdot(\cA_{y,y'})^{-1}\cdot\test{\reset{y}}
 \tag{\ref{item:C3}}
\end{align*}

This completes Steps 1 and 2 of our template. Hence, timed systems with data structures whose timing features include renamings can be analyzed by our approach, with a complexity blow-up that is only polynomial in the size of the input. We remark that, even in the case of timed automata without data structures, the presence of clock renamings makes the model exponentially more succinct~\cite{BC05}. That is, if we were to convert timed automata with such clock renamings to ordinary timed automata (using for instance the reduction from~\cite{BDFP04}) and then apply our technique, this would have incurred an additional exponential blowup that we avoid by using our template above.

\subsection{Extending to other problems: Model checking}

Here, we would like to check whether a system satisfies a specification. As
usual, we assume a finite set $\AP$ of atomic propositions which are used to
link the system and the specification, and
thus we will write specifications in the logic
$\LCPDL(\AP,\GammaDS)$.
For instance, if
$\mathsf{req},\mathsf{grant}\in\AP$,
the formula
$\forallnodes(\mathsf{req}\implies\existspath{\rightarrow^{+}}\mathsf{grant})$
says that every request should eventually be granted.
As another example, the formula
$\forallnodes((a\wedge\existspath{{\rightarrow}\cdot{\Edge{d}}})\implies
\existspath{{\rightarrow}\cdot{\Edge{d}}\cdot{\rightarrow}}a)$ says that, if
some property $a\in\AP$ holds before a message is sent over data structure $d$, then
$a$ still holds after the message is received.

Specifications are evaluated over $(\AP,\Gamma^{\DS})$-graphs.
Such graphs are generated by runs of the timed system. Again, we consider valid
sequences $\tau=\tau_1\cdots\tau_n$ of instructions over $\AP\cup\SigmaDSC$.
An instruction $\tau_i\subseteq\AP\cup\SigmaDSC$ defines the atomic
propositions $\tau_i\cap\AP$ which hold on the $i$\textsuperscript{th} event, together with the
set of operations $\tau_i\cap\SigmaDSC$ which are executed at the $i$\textsuperscript{th} event.
Let $G_\tau=(V,E,\lambda)$ be the $(\AP\cup\SigmaDSC,\GammaDS)$-graph
associated with $\tau$. When $\Sigma'\subseteq\Sigma$, we note $\pi_{\Sigma'}$ the projection on
$\Sigma'$: if $G=(V,E,\lambda)$ is a $(\Sigma,\Gamma)$-graph, then
$\pi_{\Sigma'}(G)=(V,E,\lambda')$, where $\lambda'(u)=\lambda(u)\cap\Sigma'$ for
all $u\in V$.

Let $\TS$ be a timed system with data structures $\DS$ and let $\spec\in\LCPDL(\AP,\GammaDS)$ be a specification.
 Recall that, in Theorem \ref{thm:timed}, we
 define the formula $\Psi_{\TS}=\exists p_1,\ldots,p_n~\Psi'_\TS$.
 Consider $\Psi=\exists
 p_1,\ldots,p_n~(\Psi'_\TS\wedge\neg\spec)$. Let $G=(V,E)$ be an
 $(\emptyset,\Gamma^{\DS})$-graph.
 By Theorem~\ref{thm:timed}, if
 $G\models\Psi$ then $G_{\tau}\models\Psi$ and there exists a feasible $\tau\in
 L(\TS)$ such that $G=\pi_\emptyset(G_\tau)$. Then $G_\tau\models\neg\spec$,
and since the specification uses $\AP$ only, we
 deduce that $\pi_\AP(G_\tau)\models\neg\spec$. Thus,
 as a corollary of Theorem \ref{thm:timed},
 we can construct a formula
 $\Psi\in\pdl(\emptyset,\Gamma^\DS)$ which is satisfiable over
 $(\emptyset,\Gamma^{\DS})$-linear graphs iff there is a run of the system
 which violates the specification $\spec$.

\begin{corollary}\label{thm:modelchecking}
Let $\TS$ be a timed system with data structures $\DS$ and let $\spec\in\LCPDL(\AP,\GammaDS)$ be a specification.
 We can construct a formula $\Psi$ such that,
 for all $(\emptyset,\Gamma^{\DS})$-linear graphs $G$, 
 $G\models \Psi$ iff
 there exists a \emph{feasible} $\tau\in L(\TS)$ such that
 $G=\pi_\emptyset(G_\tau)$ and $\pi_\AP(G_\tau)\not\models\spec$. The
 size of $\Psi$ is polynomial in the size of $\TS$ and $\spec$, and its intersection width is 2.
\end{corollary}

\section{Conclusion}
We studied timed systems via their behaviors depicted as graphs and reasoned about these graphs via logic \pdl. This gave rise to a problem of independent and basic interest: logical definability of realizability of weighted graphs. We showed that realizability
is definable in \pdl over sequential graphs but not definable, even in \mso, over non-sequential graphs. We developed a new logic based technique to analyze and model-check timed systems having a complex interplay of time and data structures. Potential future work is in generalizing this approach to handle a larger class of timed systems. In light of the negative result for non-sequential systems, an intriguing question is to come up with classes of concurrent systems that can be analyzed. Finally, it is worthwhile exploring if this technique can be applied in practice, in building tools for timed systems.

\bibliography{papers}

\newpage
\appendix

\section{Details in proof of Theorem~\ref{thm:timed}}
\label{app:proofoftheorem}

In this appendix, we present Lemma \ref{lem:system}, which completes the proof of Theorem~\ref{thm:timed}. Though this is similar in spirit to Theorem~\ref{thm:untimed}, for the sake of completeness and to obtain the precise complexity, we detail the construction below.

\begin{lemma}\label{lem:system}
 Let $\TS$ be a timed system with data structures whose regular language of
 sequences of timed instructions is defined by the automaton
 $\cA=(Q,\iota,F,\Delta)$.  Let $\Sigma=\Sigma^\DS_\clocks\cup
 Q=\{q_1,\ldots,q_n\}$.  We can construct a formula $\xi=\exists q_1\ldots q_n
 \xi'$, with $\xi'\in \LCPDL(\Sigma,\Gamma^\DS)$, such that, for all
 $(\emptyset,\Gamma^\DS)$ linear graphs $G$, $G\models \xi$ iff $G=\pi(G_\tau)$
 for some sequence $\tau\in L(\TS)$. Formula $\xi$ is polynomial in the size of 
 $\TS$.
\end{lemma}

\begin{proof}
 Let $\cA=(Q,\iota,F,\Delta)$ with $Q$
 the set of states, $\iota$ the initial state, $F$ the set of final states, and
 $\Delta\subseteq Q\times 2^{\Sigma^\DS_\clocks} \times Q$ the transition
 relation. We denote a transtion $\delta\in \Delta$ by
 $\delta=(\src(\delta),\lab(\delta),\tgt(\delta))$, where $\src,\tgt$ are
 source and target states and $\lab(\delta)$ is the set of timed instructions
 labeling a transition. For simplicity, we assume that the data structure
 value (e.g., messages, stack symbols, etc) is a singleton set. Later we
 indicate how this can also be extended to a finite alphabet of data structure
 values.

 Now, we write a formula $\xi$ to capture the accepted sequences of sets of
 instructions in $\EQLCPDL$. We start by guessing the states and instructions
 visited along an accepting run, i.e., we write
 \[\xi=\exists q_1\ldots q_n \xi'\]
 where $\xi'$ is built as a conjunction of the following subformulae which
 check the conditions of being an accepting run. Let us describe each
 subformula along with the property that it is expected to capture.
 \begin{itemize}
 \item \emph{States}. Every position in the run is labeled by a unique state. 
 \[\Psi_{state}=\forallnodes \bigvee_{q\in Q}\Big({q}\wedge \bigwedge_{q'\in 
 Q\setminus\{q\}} \hspace{-2mm} \neg {q'}\Big)\]
 
 \item \emph{Transitions}. Every forward edge must have a corresponding transition either from the previous source state to a next target state, or from an initial state to the target state. Further, all node labels (instructions) are those mentioned in the transitions associated with the nodes.
 \begin{align*}
 \Psi_{trans}= \forallnodes \existspath{\Edgem{}} &\implies 
 \bigvee_{\delta\in\Delta} {\tgt(\delta)} \wedge \existspath{\Edgem{}}{\src(\delta)}
 \wedge \bigwedge_{r\in\lab(\delta)} {r} \wedge 
 \bigwedge_{r\in \Sigma^\DS_\Clocks\setminus \lab(\delta)} \hspace{-5mm} \neg {r}
 \\
 {}\wedge\forallnodes \neg \existspath{\Edgem{}} &\implies 
 \bigvee_{\delta\in\Delta\mid\src(\delta)=\iota} \hspace{-4mm} \tgt(\delta) \wedge 
 \bigwedge_{r\in\lab(\delta)} {r} \wedge
 \bigwedge_{r\in \Sigma^\DS_\Clocks \setminus \lab(\delta)} \hspace{-5mm} \neg {r}
 \\
 {}\wedge\forallnodes \neg \existspath{\Edge{}} &\implies 
 \bigvee_{q\in F} q
 \end{align*}
 
 \item \emph{Data structures}. All data structure policies must be followed
 accurately. Let $\mathsf{Stacks}$ be the set of LIFO-Stacks and $\mathsf{Queues}$ be the set of
 FIFO-queues in $\DS$. Then we need to check the following properties: (i) every
 stack is LIFO; (ii) every queue is FIFO; (iii) data structure edges must go
 forward with respect to the linear order; (iv) at any transition labeled ``write'' (i.e., every node
 labeled $w(d)$), there is a unique outgoing data structure edge, and at every read
 transition (i.e., every node labeled $r(d)$), there is a unique incoming data structure edge. We
 now write the formula to capture all of these as a conjunct of formulae, each
 capturing the above properties.
 \begin{align*} 
 \Psi_{\DS} & = \bigwedge_{d\in \mathsf{Stacks}} \hspace{-2mm}
 \forallnodes(w(d) \Rightarrow
 \existsloop{{\Edge{}} \cdot ({\Edge{d}} \cdot {\Edge{}} +
 \test{\neg(w(d)\vee r(d))} \cdot {\Edge{}})^\ast \cdot {\Edgem{d}}}) \\
 & \wedge 
 \bigwedge_{d\in \mathsf{Queues}} \hspace{-2mm} \neg\existsnode\existsloop{\Edge{}^+ \cdot \Edge{d} \cdot \Edge{}^+ \cdot \Edgem{d}} \\
 & \wedge \bigwedge_{d\in \DS} \neg \existsnode \existsloop{(\Edge{}^+\cdot \Edge{d}) + (\Edge{}^+ \cdot \Edgem{d} \cdot \Edge{d}) + (\Edge{}^+ \cdot \Edge{d} \cdot \Edgem{d})}
 \\
 &\wedge \bigwedge_{d\in \DS} 
 \forallnodes ({w(d)}\Leftrightarrow \existspath{\Edge{d}}) \wedge 
 \forallnodes ({r(d)}\Leftrightarrow \existspath{\Edgem{d}}).
 \end{align*}
 For simplicity, we also make some assumptions about the data structure access -- in particular, at any node (event), only one data structure access operation is performed. Hence we cannot have a push and pop at the same time, etc. These can easily be captured as a conjunction of the below formula with $\Psi_{\DS}$.
 \begin{align*}
 \Psi_{DS}'=\bigwedge_{d\in\DS} 
 \neg\existsnode(w(d)\wedge r(d)) &
 \wedge\bigwedge_{d'\in\DS\setminus\{d\}} \hspace{-3mm} 
 \neg\existsnode((w(d)\vee r(d))\wedge(w(d')\vee r(d'))) \,.
 \end{align*}
\end{itemize}
 Define $\xi'=\Psi_{state}\wedge \Psi_{trans}\wedge \Psi_{\DS}\wedge
 \Psi_{\DS}'$. The correctness of the above formula, in encoding an accepting
 run, can be argued as follows: for any $(\emptyset, \gamma^\DS)$-labeled linear
 graph $G$ with $n$ nodes, $G\models \xi$ iff there exists a sequence of
 transitions $\delta_1,\ldots,\delta_n$ such that $\src(\delta_1)=\iota$ is the
 initial state, $\tgt(\delta_n)$ is a final state,
 $\tgt(\delta_j)=\src(\delta_{j+1})$ for all $1\leq j<n$, all data structure
 policies are followed, and the labels along this accepting run define a
 sequence $\tau\in L(\TS)$ such that $G=G_\tau$. We observe at this point that
 in the above formula we did not check whether the graph produced by the system
 was a linear graph. Indeed, this is not possible in \pdl. However, our
 statement is only about linear graphs, in other words, we assume that all
 graphs generated by our system are linear (which is of course true for any
 sequential system) and thus our proof is complete.

 To handle data structures over an arbitrary message alphabet, we simply enhance
 our propositions with the alphabet of messages $\mathsf{Msg}$. Then we can write
 formulae to check that each read or write is associated with a single message 
 and the message at a read event is the message that was written at the 
 matching write event. 
 \begin{align*} 
 \Psi_{\mathsf{msg}}&=\bigwedge_{d\in\DS} \forallnodes \Big( w(d) \implies
 \bigvee_{m\in \mathsf{Msg}} m \wedge \existspath{\Edge{}}m \Big)
 \\
 &\wedge \bigwedge_{m\in \mathsf{Msg}} \forallnodes \Big( m \implies
 \bigwedge_{m'\in \mathsf{Msg}\setminus\{m\}} \hspace{-5mm} \neg m' 
 \wedge \bigvee_{d\in\DS} w(d)\vee r(d) \Big)
 \end{align*}
 This concludes the proof.
\end{proof}

\end{document}

%% file: preamble.tex
\usepackage{url}
\usepackage{booktabs}
\usepackage{subcaption}

\usepackage{stackrel,hyperref}
\usepackage{balance}
\usepackage{breakurl}             
\usepackage{wrapfig}
\usepackage{verbatim}
\usepackage{color}
\usepackage{xspace}
\usepackage{epsfig}
\usepackage{savesym}
\usepackage{amsmath,amssymb}
\usepackage{wasysym}
\usepackage{stmaryrd}
\usepackage{enumitem}

\usepackage{thm-restate}
\usepackage{pgf}

\usepackage[colorinlistoftodos,prependcaption,textwidth=3.5em,textsize=small]{todonotes}

\usepackage{tikz}
\usetikzlibrary{fit} 
\usetikzlibrary{backgrounds} 
\usetikzlibrary{patterns,calc}
\usetikzlibrary{decorations.pathreplacing}
\usetikzlibrary{positioning}
\usetikzlibrary{decorations.pathmorphing}
\usetikzlibrary{decorations.markings}
\usepgflibrary{shapes}
\usetikzlibrary{snakes,automata,arrows}
\usetikzlibrary{shadows}
\tikzstyle{background}=[rectangle,fill=gray!10, inner sep=0.1cm, rounded corners=0mm]
\tikzstyle{vecArrow} = [decoration={markings,mark=at position 1 with 
     {\arrow[semithick,out=50,in=20]{triangle 60}}}, shorten >= 5.5pt, ]
\tikzstyle{cir}=[draw=violet, fill=violet!20!white, rounded rectangle,minimum size=1.5em,inner sep=0em]                                                                                                      
     \tikzstyle{cir1}=[draw=green!50!blue, fill=green!40!white, circle,minimum size=1.5em,inner sep=0.1em]                                                                                                      
\tikzstyle{cir2}=[draw=red,fill=red!20!white,rounded rectangle,minimum size=1.5em,inner sep=0em]                                                                                                    
\tikzstyle{cir11}=[draw=green!20!blue,fill=green!40!white,rounded rectangle,minimum size=1.5em,inner sep=0em]                                                                                                    
\tikzstyle{cir22}=[draw=green!70!violet,fill=green!15!violet!40!white,rounded rectangle,minimum size=1.5em,inner sep=0em]

\newcommand{\dom}{\mathsf{dom}}

\DeclareMathOperator{\bow}{\bowtie}

\newcommand{\Succ}{\mathsf{succ}}

\newcommand{\edg}[1]{\mathrel{E}_{#1}}

\newcommand{\po}{\preceq}
\newcommand\podot{\mathrel{\ooalign{$\prec$\cr
  \hidewidth\raise-0.025ex\hbox{$\cdot\mkern-0.5mu$}\cr}}}
\newcommand{\proc}{\rightarrow}

\newcommand{\ts}{\mathsf{ts}}
\newcommand{\tsm}{\mathsf{tsm}}
\newcommand{\src}{\mathsf{src}}
\newcommand{\tgt}{\mathsf{tgt}}

\newcommand{\dtsm}{\mathsf{d_{tsm}}}
\newcommand{\dtsmp}{\mathsf{d_{tsm}^+}}

\newcommand{\Gg}{\mathcal{G}}
\newcommand{\Ss}{\mathcal{S}}
\newcommand{\Tt}{\mathcal{T}}
\newcommand{\lab}{\mathsf{lab}}
\newcommand{\tq}{:}

\newcommand{\mso}{\ensuremath{\mathsf{MSO}}\xspace}
\newcommand{\pdl}{\ensuremath{\mathsf{EQ\text{-}ICPDL}}\xspace}
\newcommand{\existsnode}{\mathsf{E}\,}
\newcommand{\forallnodes}{\mathsf{A}\,}
\newcommand{\existspath}[1]{\ensuremath{\langle#1\rangle}}

\newcommand{\existsloop}[1]{\mathsf{loop}(#1)}
\newcommand{\test}[1]{\mathsf{test}\{#1\}}

\newcommand{\intp}[1]{\lfloor{{#1}\rfloor}}
\newcommand{\fracp}[1]{\lbrace{#1}\rbrace}
\newcommand{\pos}{\prec}
\newcommand{\lef}{\mathsf{geq_{Fr}}}
\newcommand{\ltf}{\mathsf{gt_{Fr}}}
\newcommand{\ford}{\lef}
\newcommand{\fords}{\ltf}

\newcommand{\cS}{\mathcal{S}}

\newcommand{\cA}{\mathcal{A}}

\newcommand{\TCG}{\mathsf{TCGraphs_M}}

\newcommand{\DS}{\mathsf{DS}}

\newcommand{\nxt}{\mathsf{next}}

\newcommand{\nop}{\mathrm{nop}}

\newcommand{\pop}{\mathsf{pop}}
\newcommand{\push}{\mathsf{push}}

\newcommand{\reset}[1]{(#1:=0)} 
\newcommand{\sem}[1]{\llbracket {#1} \rrbracket}

\newcommand{\poly}{\textsf{poly}}

\newcommand{\TS}{\mathcal{T}}
\newcommand{\wg}{\mathcal{G}}

\newcommand{\Graphs}{\mathcal{G}}
\newcommand{\AP}{\mathsf{AP}}
\newcommand{\PDL}{\mathsf{PDL}}
\newcommand{\CPDL}{\mathsf{CPDL}}
\newcommand{\LCPDL}{\mathsf{LCPDL}}
\newcommand{\ICPDL}{\mathsf{ICPDL}}
\newcommand{\EQLCPDL}{\ensuremath{\mathsf{EQ\text{-}LCPDL}}\xspace}
\newcommand{\EQICPDL}{\ensuremath{\mathsf{EQ\text{-}ICPDL}}\xspace}
\newcommand{\lleq}{\mathrel{\triangleleft}}
\newcommand{\clocks}{\mathsf{Clocks}}
\newcommand{\Clocks}{\clocks}
\newcommand{\Edge}[1]{\xrightarrow{#1}}
\newcommand{\Edgem}[1]{\mathrel{{\xrightarrow{#1}}\strut^{-1}}}
\newcommand{\creset}{\mathsf{Reset}}

\newcommand{\spec}{\Phi}
\newcommand{\myst}{\,:\,} 

 \DeclareMathOperator{\fole}{\textcolor{blue!60!green}{4<}}
 \DeclareMathOperator{\leqfi}{\textcolor{red!50!yellow}{\leq 5}}